\documentclass[prd,nofootinbib,preprint,superscriptaddress]{revtex4}

\usepackage{amsmath, amssymb, amsthm, graphicx, epsfig, fancyhdr,epsfig, slashed, mathrsfs,xstring,amsfonts,bbold}

\usepackage{tikzsymbols}
\usepackage{tikz,xcolor}
\usepackage{natbib}
\usepackage{float}

\usepackage{mathtools}

\usepackage[colorlinks = true,linkcolor = cyan(process),urlcolor  = magenta,citecolor = red,anchorcolor = blue]{hyperref}

\definecolor{cyan(process)}{rgb}{0.0, 0.72, 0.92}

\definecolor{lime}{HTML}{A6CE39}
\DeclareRobustCommand{\orcidicon}{
	\begin{tikzpicture}
	\draw[lime, fill=lime] (0,0) 
	circle [radius=0.2] 
	node[white] {{\fontfamily{qag}\selectfont \tiny ID}};
	\draw[white, fill=white] (-0.0625,0.095) 
	circle [radius=0.007];
	\end{tikzpicture}
	\hspace{-2mm}
}

\foreach \x in {A, ..., Z}{\expandafter\xdef\csname orcid\x\endcsname{\noexpand\href{https://orcid.org/\csname orcidauthor\x\endcsname}
			{\noexpand\orcidicon}}
}

\newcommand{\be}{\begin{equation}}
\newcommand{\ee}{\end{equation}}
\newcommand{\bea}{\begin{eqnarray}}
\newcommand{\eea}{\end{eqnarray}}

 \usepackage{ulem,soul,multirow,upgreek,cancel,physics}
 \usepackage{empheq}
\usepackage[subnum]{cases}

\usepackage{xpatch}
\makeatletter
\xpatchcmd{\@ssect@ltx}{\@xsect}{\protected@edef\@currentlabelname{#8}\@xsect}{}{}
\xpatchcmd{\@sect@ltx}{\@xsect}{\protected@edef\@currentlabelname{#8}\@xsect}{}{}
\makeatother
\makeatletter
\newcommand\colorwave[1][blue]{\bgroup \markoverwith{\lower3.5\p@\hbox{\sixly \textcolor{#1}{\char58}}}\ULon}
\font\sixly=lasy6 
\makeatother

\newcommand{\eq}[1]{\begin{align}#1\end{align}}

\newcommand{\ba}{\begin{eqnarray}}
\newcommand{\ea}{\end{eqnarray}}

\newcommand{\bad}{\begin{aligned}}
\newcommand{\ead}{\end{aligned}}

\newcommand{\lt}{\left }
\newcommand{\rt}{\right }

\newcommand{\gsim}{\gtrsim}
\newcommand{\lsim}{\lesssim}

\newcommand{\unit}[1]{\, \text{#1}}

\newcommand{\tb}[1]{\, \textbf{#1}}

\def\mpl{M_P}

\def\cs{{a}}

\def\dr{\text{DR}}
\def\cmb{\text{CMB}}
\def\bbn{\text{BBN}}

\def\GeV{\unit{GeV}}

\newcommand{\Planck}{\textit{Planck}}

\newcommand{\KeckArray}{\textit{Keck Array}}
\newcommand{\BICEP}{\textsc{Bicep}}

\newcommand{\CMBBharat}{\text{CMB-Bh$\overline{a}$rat}}
\def\sptnew{\textsc{spt-3g}}

\newcommand{\cmbsfour}{{CMB-S4}}

\newcommand{\coremfive}{\textit{\negthinspace CORE\/}}

\newcommand{\Neff}{\ensuremath{N_\mathrm{eff}}}
\newcommand{\neff}{\Neff}
\def\Trh{T_{\rm RH}}
\def\ncmb{{\cal N}_{\cmb}}
\def\nrh{{\cal N}_{\rm RH}}
\def\gsrh{g_{\star{,\rm RH}}}
\def\gssrh{g_{\star,s{,\rm RH}}}
\def\rr{\r_{\rm rad}}

\def\sm{\text{SM}}
\def\bsm{\text{BSM}}

\def\td{\text{d}}

\def\D{\Delta}
\def\g{\gamma}
\def\ug{\upgamma}
\def\n{\nu}
\def\r{\rho}
\def\a{\alpha}
\def\b{\beta}
\def\m{\mu}
\def\G{\Gamma}
\def\L{\Lambda}
\def\s{\sigma}
\def\vp{\varphi}

\def\mphi{m_\phi}
\def\lpx{\lambda_{\phi X} }
\def\lh{\lambda_{12, H}}
\def\bx{B_X}

\def\hs{\overline{H}}

\def\pcc{\phi \to \tilde{\chi}\chi}

\newcommand{\cmbs}{0.06}

\usepackage[capitalise]{cleveref}
\Crefname{figure}{Fig.}{Figs.}
\Crefname{section}{Sec.}{Secs.}
\newcommand{\Ccite}[1]{%
\IfSubStr{#1}{,}{Refs.~}{Ref.~}\cite{#1}%
}

\begin{document}


\title{Measuring inflaton couplings via dark radiation as $\Delta N_{\rm eff}$ in CMB}

\author{Anish Ghoshal\orcidA{}}
\email{anish.ghoshal@fuw.edu.pl}
\author{Zygmunt Lalak,\orcidC{}}
\email{zygmunt.lalak@fuw.edu.pl}
\affiliation{Institute of Theoretical Physics, Faculty of Physics, University of Warsaw, ul. Pasteura 5, 02-093 Warsaw, Poland}

\author{Shiladitya Porey\orcidB{}}
\email{shiladityamailbox@gmail.com}
\affiliation{Department of Physics, Novosibirsk State University, Pirogova 2, 630090 Novosibirsk, Russia}

\begin{abstract}
\textit{We study the production of a beyond the Standard Model (\bsm) free-streaming relativistic particles which contribute to $N_{eff}$ and investigate how much the predictions for the inflationary analysis change. We consider inflaton decay as the source of this dark radiation (\dr) and use the Cosmic Microwave Background (\cmb) data from \Planck-2018 to constrain the scenarios and identify the parameter space involving couplings and masses of the inflaton that will be within the reach of next-generation CMB experiments like \sptnew, \cmbsfour, \CMBBharat, PICO, CMB-HD, etc. We find that if the \bsm~particle is produced only from the interaction with inflaton along with Standard Model (\sm)~relativistic particles, then its contribution to $\neff$ is a monotonically increasing function of the branching fraction, $\bx$ of the inflaton to the \bsm~particle $X$; \Planck~bound on $\neff$ rules out such $\bx\gsim 0.09$. 
Considering two different analyses of \Planck +\BICEP~data together with other cosmological observations, $\neff$ is treated as a free parameter, which relaxes the constraints on scalar spectral index ($n_s$) and tensor-to-scalar ratio ($r$). 
The first analysis leads to predictions on the inflationary models like Hilltop inflation being consistent with the data. Second analysis rules out the possibility that \bsm~particle $X$ producing from the inflaton decay in Coleman-Weinberg Inflation or Starobinsky Inflation scenarios. To this end, we assume that \sm~Higgs is produced along with the \bsm~particle. We explore the possibilities that $X$ can be either a scalar or a fermion or a gauge boson and consider possible interactions with inflaton and find out the permissible range on the allowed parameter space Planck and those which will be within the reaches of future \cmb~observations.
}
\end{abstract}

\maketitle

\section{Introduction}

For resolving the horizon problem, the flatness problem and lay the seed for structure formation in the late universe cosmic inflation~\cite{Starobinsky:1980te, Sato:1980yn, Guth:1980zm, Linde:1981mu, Albrecht:1982wi, Linde:1983gd} is the leading paradigm. This period in the very early universe involves an accelerated expansion epoch during which vacuum quantum fluctuations of the gravitational and matter fields were amplified to large-scale cosmological perturbations~\cite{Starobinsky:1979ty, Mukhanov:1981xt, Hawking:1982cz,  Starobinsky:1982ee, Guth:1982ec, Bardeen:1983qw}, that later seeded the anisotropies as observed in Cosmic Microwave Background Radiation (CMBR) and lead to the formation of the Large Scale Structure (LSS) of our Universe.

What we observe in CMBR can be accounted for in a minimal setup, where inflation is driven typically by a single scalar field $\phi$ with canonical kinetic term, minimally coupled to gravity, and evolving in a flat potential $V(\phi)$ in the slow-roll regime. Since particle physics beyond the electroweak scale remains elusive and given that inflation can proceed at energy scales as large as $10^{16}\GeV$, even within this class of models, hundreds of inflationary scenarios have been proposed to match with the latest sophisticated measurements of CMB \cite{Ade:2013sjv, Adam:2015rua, Array:2015xqh, Ade:2015lrj}. A systematic Bayesian analysis reveals that one-third of them can now be considered as ruled out~\cite{Martin:2013tda, Martin:2013nzq, Vennin:2015eaa}, while the vast majority of the preferred scenarios are of the plateau type, i.e., they are such that the potential $V(\phi)$ is a monotonic function that asymptotes a constant value when $\phi$ goes to infinity.


Cosmic inflation must be succeeded by a period during which the inflaton's energy must be transferred to relativistic \sm~particles, resulting in the formation of a hot universe consistent with present observations. The extremely adiabatic reheating epoch is important since it produces all matter in the cosmos as well as the relativistic \sm~fluid that raises the temperature of the universe. During this epoch, the standard picture is that the inflaton field vibrates coherently around the minimum of its potential, producing \sm~particles as well as viable \bsm~particles via gravitational interactions or other couplings that maybe available in the theory. 
This \bsm~particle may contribute to the dark matter sector or stay relativistic, contributing to dark radiation (\dr). If they contribute to \dr, they may influence the expansion rate of the universe (and thus the measured value of Hubble parameter), CMB anisotropy~\cite{Menegoni:2012tq}, and the perturbation for large scale structure-formation in the universe~\cite{Archidiacono:2011gq}. Moreover, \dr~can also be produced from the decays of massive particles~\cite{Hasenkamp:2012ii}, $U(1)$ gauge field which is singlet under \sm~\cite{Ackerman:2008kmp}. There are several proposed particles as viable candidate from \dr~e.g., light sterile neutrino~\cite{Archidiacono:2016kkh, Archidiacono:2014apa}, neutralino~\cite{Bae:2013hma}, axions~\cite{DiValentino:2013qma,DEramo:2018vss}, Goldstone
bosons~\cite{Weinberg:2013kea}, early dark energy~\cite{Calabrese:2011hg}.

The predictions of inflationary models, such as $n_s$ and $r$, are solely determined by the parameters of the models and are independent of the presence of \dr~in the universe. However, the presence of extra relativistic species in the standard cosmology does have an impact on the selection of inflationary models. In particular, single-field slow-roll models, which are known to produce high values of the scalar spectral index in standard cosmology may be more favorable in the presence of extra relativistic species.  Upcoming \cmb~experiments e.g. \sptnew, CMB Stage IV (\cmbsfour), \CMBBharat, PICO, and CMB-HD, all of which are highly sensitive to the additional relativistic degrees of freedom, are expected to offer  more precise information on the extra relativistic \bsm~particles present along with \cmb~photons. Gaining this information is significant for developing a more complete and accurate theory of the physics of the early universe, including inflationary epoch followed by reheating era, and has important impacts on understanding the underlying physics of the thermal history, Hubble expansion rate, and other cosmological phenomena of the later universe.

In this work, we make the assumption that the \dr~is created during the reheating epoch as a relativistic non-thermal \bsm~particle together with relativistic \sm~particles, and that the inflaton is the only source of this radiation. We then apply the bounds on \dr~from the \cmb~to the branching fraction for the production of extra relativistic \bsm~particle. Next, we explore the possibilities of the extra \dr~particle being a fermion, boson, or gauge boson and consider the possible leading-order interactions. Utilizing the present bound from current \cmb~observations and prospective sensitivity reaches on \dr~from future \cmb~experiments, we proceed to identify the parameter space that involves the inflaton mass, as well as the couplings between the inflaton and both the visible sector and the \dr~particle.

 
\textit{The paper is organized as follows:} we begin with a discussion of $\neff$ and $\D\neff$ in Section~\ref{Sec:Effective number of  relativistic degrees of freedom}. In Section~\ref{Sec:Dark Radiation and Inflation}, we 
consider $n_s-r$ predictions from four disparate inflationary models and review whether they can be ruled out as viable inflationary models by the bound from \Planck2015+\BICEP2 data while $\neff$ is allowed to vary from its standard value. In Section~\ref{Sec:Inflaton decay during reheating}, we consider the production of a \bsm~particle during the reheating era, which adds an extra relativistic degree to \cmb. We also explore the connection between branching fraction for the production of that \bsm~particle and $\D\neff$. In Section~\ref{Sec:Inflaton Decay}, we look at probable interactions between the inflaton and the \bsm~particle, as well as the permitted parameter space for the couplings related to the bound on $\neff$ from \cmb. In Section~\ref{Sec:Discussion and Conclusion}, we summarize the findings.

In this work, we use the natural unit with $\hbar=c=k_{B}=1$, such that the reduced Planck mass is $M_P\simeq 2.4 \times 10^{18}\, \text{GeV}$. Furthermore, we also assume that the signature of the space-time metric is $(+,-,-,-)$.

\section{Effective number of relativistic degrees of freedom}
\label{Sec:Effective number of  relativistic degrees of freedom}
Since the temperature of the \cmb~photons is a well-known quantity, the current total energy density of the relativistic species of the universe, $\r_{{\rm rad},{\rm tot}} $, can be expressed in terms of the energy density of \sm~photon, $\r_\g$, as~\cite{Archidiacono:2011gq,Mangano:2001iu,ParticleDataGroup:2020ssz,Boehm:2012gr,Paul:2018njm,Hasenkamp:2012ii,Tram:2016rcw}
\eq{
\r_{{\rm rad},{\rm tot}} &=  \r_\g \qty[ 1+ \frac78 \lt(\frac{4}{11}\rt)^{4/3}  \Neff ]\,,
\label{Eq:Neff-def}
}
where $\neff$, also known as {\it effective number of relativistic  degrees of freedom}~\cite{Archidiacono:2011gq}, parameterizes the contribution from non-photon relativistic particles, such as \sm~cosmic neutrinos. 
There are three left-handed light cosmic neutrinos, and they were in thermal equilibrium with the \sm~relativistic particles in the hot early universe. When the temperature of the universe drops to $800\unit{keV}$, neutrinos decouple from the \sm~photons just before the electron-positron annihilation. If we assume neutrinos decouple instantly, their contribution to $\neff$ is expected to be $3$. However, if we consider noninstantaneous decoupling of cosmic neutrinos~\cite{Husdal:2016haj}, QED correction~\cite{Mangano:2001iu}, and three-neutrino flavor oscillations on the neutrino decoupling phase~\cite{Mangano:2005cc} (see also \cite{Bennett:2019ewm}), then, neutrinos get partially heated during electron-positron decoupling from photon, resulting in slightly higher temperature of neutrinos.~
This leads to~\cite{EscuderoAbenza:2020cmq} (see also~\cite{deSalas:2016ztq})
\footnote{
Non-instantaneous decoupling of neutrinos from $e^- e^+$ pairs during the early universe leads to an increase in the number of equilibrium neutrino species by $\D N_\n = 0.035$~\cite{Dolgov:2002wy,Bambi:2015mba,Dolgov:1992iw,Dolgov:1992qg}. The effective number of neutrino species receives an additional contribution $\D N_\n = 0.011$ resulting from the deviation of the electron-positron-photon plasma from an ideal gas state. For further details, see~\Ccite{Dolgov:2002wy,Bambi:2015mba}. Consequently, within the framework of the \sm, the effective neutrino species amounts to ${\Neff}_{,\sm} = 3.046$.}
\eq{\label{Eq:3.046}
{\Neff}_{,\sm} = 3.046\,.
}
Any measured value higher than Eq.~\eqref{Eq:3.046} suggests the possibility of the existence of any relativistic \bsm~particle%
\footnote{
If the \bsm~particle is thermal, with a higher value of couplings with electron-positron than with neutrinos, it may result in $\neff<{\Neff}_{,\sm}$~\cite{Ho:2012ug,RoyChoudhury:2022rva}.
}%
. The contribution of the \bsm~particle in $\Neff$ is expressed as~\cite{ParticleDataGroup:2020ssz,Abazajian:2019eic,Abazajian:2019oqj,Luo:2020sho}
\eq{\label{Eq:Definition-of-DNeff}
\D \Neff&=\Neff - {\Neff}_{,\sm}\,.
}
Current bounds on and prospective future sensitivity reaches on $\D\Neff$ from that will be within the reach of future \cmb~measurements are mentioned in Table~\ref{Table:bound-on-Delta-Neff}. 
%
%
\begin{table}[htp!]
\begin{center}
\caption{\it Bounds on $\Neff$ (or $\D\Neff$) from present \cmb~observations and prospective future reaches of $\Neff$ (or $\D\Neff$) that upcoming \cmb~experiments may be able to observe~\cite{Berbig:2023yyy, Gerbino:2022nvz}.
}
\label{Table:bound-on-Delta-Neff}
\begin{tabular}{ |c| c|  }
\hline
 $\D {\Neff}_{\Planck} =0.29\, 
 $ & \cite{Planck:2018vyg}
 \\
 \hline
  $\D {\Neff}_{\Planck + BAO} =0.28\,
  $ &
  \cite{Planck:2018vyg}\\
  \hline
 $\D {\Neff}_{\sptnew} = 0.2$ & 
 \cite{SPT-3G:2021wgf}\\
 \hline
 $\D {\Neff}_{\coremfive+BAO} = 0.039$ 
 & \cite{CORE:2017oje}\\
 \hline
 $\D {\Neff}_{\CMBBharat} =\lt(3.05^{+0.05}_{-0.07} -3.046\rt)$ &  \cite{CMBBharat:01}\\
 \hline
  $\D {\Neff}_{PICO} 
  =0.06$ at $2\sigma$ & \cite{NASAPICO:2019thw} \\
   \hline
 $\D {\Neff}_{\cmbsfour} = \cmbs$ 
 at $95\%$ CL
 & \cite{CMB-S4:2022ght,Abazajian:2019oqj}\\
  \hline
    $\D {\Neff}_{CMB-HD} \lsim 0.027 $ & \cite{Sehgal:2019ewc,CMB-HD:2022bsz,Cheek:2022dbx} \\
  \hline
\end{tabular}
\end{center}
\end{table}
%
%
In this work, we assume that a relativistic non-thermal non-self-interacting \bsm~particle contributes to $\r_\g$ which is produced from the inflaton during reheating era.

\section{Inflation and Dark Radiation}
\label{Sec:Dark Radiation and Inflation}
%
%

\begin{figure}[htp!]
    \centering
    \includegraphics[width=0.75\linewidth]
    {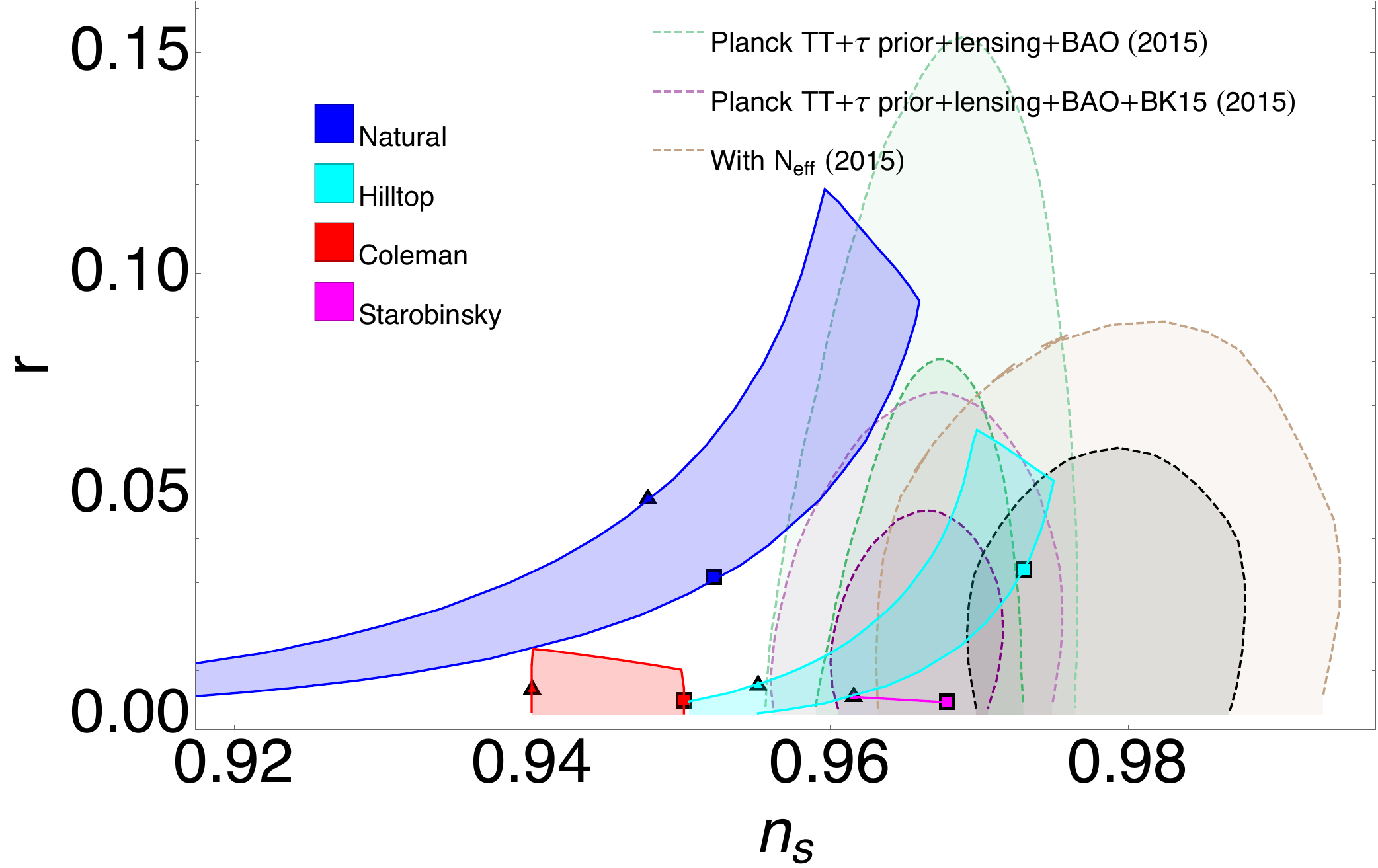}
    \caption{\it \raggedright \label{Fig:Planck15+BK15-inflation-ns-r-bound} 
    $n_s-r$ predictions for four inflationary models (Starobinsky Inflation, Natural Inflation, Hilltop Inflation, and Coleman-Weinberg Inflation)
    ~for $50 \lsim N_{\rm CMB} \lsim 60$. The curves with a triangle on them correspond to $N_{\rm CMB}=50$, while the curves with a square on them correspond to $N_{\rm CMB}=60$.  The green-shaded regions indicate the bound on $(n_s, r)$ plane at $68\%$ and $95\%$ CL from the \Planck2015~\cite{Planck:2015fie}, and the purple-colored regions show bounds from the combined analysis of \Planck2015+\BICEP2-\KeckArray2015~\cite{BICEP2:2018kqh} for a standard value of 
    %
    %
    %
    %
    $\D\Neff$ ($\D\Neff=0$). The regions with brown shading show the shifting of the same bounds when $\D\Neff$ is varied (using \Planck2015+\BICEP2-\KeckArray2015+baryon acoustic oscillation data + direct measured value of present-day Hubble constant, with $\Lambda$CDM+$r+\Neff$ as parameters lead $\D\neff=0.254$)~\cite{Guo:2017qjt,Kannike:2018zwn}.  Predictions from Hilltop and Starobinsky inflation are within $1-\sigma$ CL from \Planck2015+\BICEP2-\KeckArray2015 bound. However, when $\D\Neff$ is used as a free parameter, S-I survives at $2-\sigma$ bound for higher number of $\ncmb$ while H-I can satisfy $(n_s,r)$ values within 
    $1-\sigma$ bound.
%
      }  
\end{figure}

In this section, we review how the non-zero value of $\D\neff$ influences the selection of single-field slow-roll inflationary models. If $V(\phi)$ is the potential energy of a single real scalar inflaton $\phi$, minimally coupled to gravity, then the action with canonical kinetic energy is~\cite{Senatore:2016aui}
\be 
{\cal S}\supset \frac{\mpl^2}{2} \int \td^4 x \, \sqrt{-g} \lt({\cal R} +\frac12\partial_\m\phi\partial^\n\phi - V(\phi)\rt) 
\,. \label{Eq:KG_Action}
\ee 
 Here, 
 $g$ is the determinant of the spacetime metric tensor $g_{\m\n}$, and ${\cal R}$ is the Ricci scalar. 
 For slow roll inflationary scenario driven by $\phi$, 
the first two potential-slow-roll parameters are~\cite{Liddle:1994dx} 
\begin{eqnarray}
  \epsilon_V (\phi) =\frac{\mpl^2}{2} \left(\frac{V'(\phi )}{V(\phi )}\right)^2 ,\qquad 
 && \eta_V (\phi) = \mpl^2\frac{V''(\phi )}{V(\phi )} \,,
\end{eqnarray}
where 
prime symbolizes derivative with respect to $\phi$. 
The slow-roll inflationary period is continued till $\dot{\phi}^2\ll V(\phi)$ and $\ddot{\phi}\ll 3 {\cal H}\dot{\phi}$~\cite{Ryden:1970vsj,Riotto:2002yw,Lyth:1993eu}, with over dot implying a derivative with respect to physical time $t$ and ${\cal H}\equiv {\cal H}(t)$ being the Hubble parameter, are maintained. These two conditions lead to %
$\epsilon_V (\phi), \lt|\eta_V (\phi)\rt|\ll 1$. By the time either of these two parameters becomes $\sim 1$ at $\phi=\phi_{\rm end}$, slow roll epoch ends. The duration of the inflationary epoch is parameterized by the number of $e$-folds, $\ncmb$, as,
\begin{eqnarray}
\ncmb \equiv \int_{\cs_{\rm end}}^{\cs^*} \td \log(\cs) = {1\over \mpl^2}\int_{\phi_{\rm end}}^{\phi^*} {V(\phi)\over V'(\phi)} \, \td\phi\,.
\end{eqnarray}
Here, $\phi^*$ is the value of inflaton corresponding to both cosmological scale factor $\cs^*$ and the length-scale of \cmb~observation, and $\cs_{\rm end}$ is the cosmological scale factor at the end of inflation. When the value of inflaton is $\phi^*$, the length scale corresponding to $e$-fold $\ncmb$ leaves the causal horizon during inflation~\cite{Lillepalu:2022knx}. %
The largest length scale of the \cmb~that can be observed today formed at$\sim 60$ $e$-folds before the end of inflation~\cite{Baumann:2022mni}. As $\ncmb \gsim 60$ and $\ncmb\gsim 40$ are required to solve the horizon and the flatness problems, we varied $\ncmb$ from $50$ to $60$ in this work. %
On the other hand, quantum fluctuations can be generated by the inflaton during the inflationary epoch. The statistical nature of these primordial fluctuations is expressed in terms of the power spectrum. The scalar and tensor power spectrums for '$k$'-th Fourier mode are defined as 
\ba
&&\mathcal{P}_s \left( k \right) = A_s \left(  \frac{k}{k_*} \right)^{n_s -1 + (1/2) \alpha_s \ln(k/k_*) + (1/6)\beta_s (\ln(k/k_*))^2 }  \label{eq:define scalar power spectrum}\,,\\
&& \mathcal{P}_h \left( k \right) = A_t \left(  \frac{k}{k_*} \right)^{n_t + (1/2) d n_t/d \ln k \ln(k/k_*) + \cdots } \,,\label{eq:define tensor power spectrum}
\ea
where $A_s$ and $A_t$ are the amplitudes~\cite{Racioppi:2021jai} of the respective power spectrums, $k_*$ is the pivot scale corresponding to \cmb~measurements (also corresponding to $\phi^*$ as previously mentioned), $n_s$ and $n_t$ are scalar and tensor spectral indexes, respectively. $\a_s$ and $\b_s$ are the running and running of running of scalar spectral index. The relation among $n_s$ and potential-slow-roll parameters at leading order is
\begin{eqnarray}
 n_s = 1-6\, \epsilon_V(\phi^*) + 2 \, \eta_V(\phi^*)\,.
\end{eqnarray}
On the other hand, tensor-to-scalar ratio is defined as
\begin{eqnarray}
 r =\frac{A_t}{A_s}\approx 16 \, \epsilon_V(\phi^*)\,,
\end{eqnarray}
where the last relationship is only applicable in the case of slow-roll inflationary scenario. Now, %
bounds on the $(n_s, r)$ plane at $1-\sigma$ and $2-\sigma$ CL from \Planck2015~\cite{Planck:2015fie}, and combined \Planck2015+\BICEP2-\KeckArray2015 data ($\Lambda$CDM$+r$ model)~\cite{BICEP2:2018kqh} are shown in Fig.~\ref{Fig:Planck15+BK15-inflation-ns-r-bound} (shaded contours with green and purple color, with dashed-lines as periphery). In this work, we use deep colored region (inside) for $95\%$ ($1-\s$ CL) and light colored region (outside) for $68\%$ ($2-\s$ CL) observational contours. 
In addition, 
$n_s-r$ predictions for four single-field slow-roll inflationary models~\cite{Planck:2018jri,Barenboim:2019tux}
are also shown in that figure. %
Two of these models have been disfavored at $95\%$ CL due to the smallness of the predicted value of $n_s$, while the predictions about the values of $n_s$ and $r$ from the other two models are still within the $95\%$ CL contour of \Planck2015-\BICEP2-\KeckArray2015 combined data%
\footnote{
 See~\cref{NewBound} for updated bounds.
}. 
The four inflationary models we study in this paper are quite distinct from each other. While  the Starobinsky Inflation explains the exponential expansion during inflation based on geometry without resorting to any \bsm~field, another inflationary model knwon as the Natural Inflation is based on a well-motivated \bsm~particle physics theory of axion physics which solves the strong CP problem. Furthermore, Hilltop inflationary scenario approximately resembles to any inflationary scenario that occurs near the maximum of the potential. The radiative quantum correction to the self coupling of inflaton is captured within the framework of the Coleman-Weinberg inflation.
In the following subsections, we discuss the slow roll inflationary scenario and prediction for $n_s,r$ for these four models. 

\subsection{Starobinsky Inflation ${\cal R}^2$ (S-I)}
\label{Sec:SI}
Starobinsky Inflationary (abbreviated as S-I) model  is one of the simplest examples of plateau potentials with only one parameter, where no fine-tuning%
\footnote{For sufficient inflation satisfying \cmb~data in single-field slow roll inflationary model, self-coupling and quartic coupling should be very weak. This minuscule value is referred to as the fine-tuning problem~(see Ref.~\cite{Freese:1990rb} and Ref.~\cite{Stein:2021uge} and references therein).}%
~is needed for the inception of inflation. 
This is the inflationary model which depicts inflation by adding a term proportional to the square of Ricci curvature as a modification to Einstein-gravity. 
The action of S-I in Jordan frame is given by~\cite{Starobinsky:1980te,Mazumdar:2019tbm}
\be 
{\cal S}= \frac{\mpl^2}{2} \int \td^4 x \sqrt{-g^{JF}} \lt({\cal R}^{JF} + \frac{{{\cal R}^{JF}}^2}{6 m_{SI}^2} \rt)\,.
\ee 
Here, superscript $JF$ implies that the corresponding parameter is defined in Jordan frame and $m_{SI}$ is the mass scale. This action can be converted to Einstein frame by a conformal transformation of the metric~\cite{DeFelice:2010aj} $g_{\m\n}= \Omega^2g^{JF}_{\m \n}$ where $\Omega^2 =\ln \lt( 1+ \frac{{\cal R}^{JF}}{3 m_{SI}^2} \rt)$. 
And then the action in $g_{\m \n}$ metric space takes a form similar to Eq.~\eqref{Eq:KG_Action}. Then the potential takes the form
\begin{eqnarray}\label{eq:pot-Starobinsky}
 V(\phi) = \Lambda ^4 \left(1-e^{-\sqrt{\frac{2}{3}} \frac{\phi}{\mpl} }\right)^2, \label{Eq:Pot_SI}
\end{eqnarray}
where the inflaton $\phi$ and $\Lambda^4$ are defined as~\cite{Aldabergenov:2018qhs, Martin:2013tda}
\ba 
\phi =\sqrt{\frac{3}{2}}\mpl \ln \lt( 1+ \frac{{\cal R}^{JF}}{3 m_{SI}^2} \rt)\,,  \quad  &&\Lambda^4 =\frac{3}{4}\mpl^2 \, m_{SI}^2\,. \label{Eq:mSI}
\ea 
To satisfy \cmb~data of scalar fluctuations, $m_{SI}/\mpl\sim {\cal O}(10^{-5})$. 
Now, 
the scalar spectral index and tensor-to-scalar ratio for the potential of Eq.~\eqref{Eq:Pot_SI} are 
\ba 
n_s=\frac{3\,  e^{2 \sqrt{\frac{2}{3}} \frac{\phi}{\mpl}} -14 \, e^{\sqrt{\frac{2}{3}} \frac{\phi}{\mpl}}-5}{3 \left(e^{\sqrt{\frac{2}{3}} \frac{\phi}{\mpl}}-1\right)^2} \,, \quad  \quad 
&&r= \frac{64}{3 \left(e^{\sqrt{\frac{2}{3}} \frac{\phi}{\mpl}}-1\right)^2} \,.
\ea 

The slow-roll inflationary epoch ends when $\epsilon_V\sim 1$ happens. 
Furthermore, this S-I model predicts a very small value of $r$, and thus it is within $1-\sigma$ contour of the \Planck2015+\BICEP2 \cmb~bound (see the magenta colored region (line) in Fig.~\ref{Fig:Planck15+BK15-inflation-ns-r-bound} for $50\leq N_{\rm CMB} \leq 60$. Since $n_s$ and $r$ do not depend on any parameter, this model cannot be further adjusted for the $n_s-r$ contour when $\D\Neff$ is 
treated as a variable.

\subsection{Natural Inflation (N-I)}
\label{Sec:NI}
For slow-rolling of the inflaton, the flatness of the potential needs to be maintained against radiative correction arising from self-interaction of the inflaton or its interaction with other fields. When the axion or the axionic field plays the role of inflaton, it provides the requisite mechanisms to protect the flatness of the potential. It also offers a proper explanation from particle physics, for the small values of the self-coupling and, thus, dilutes the issue of fine-tuning. That's why this inflationary model is called Natural Inflation (abbreviated as N-I). Axion is a pseudo-Nambu-Goldstone boson that arises as a result of the spontaneous breaking of global symmetry followed by additional explicit symmetry breaking~\cite{Freese:1990rb}.   
The axion potential which arises due to the spontaneous breaking of global shift symmetry or axionic symmetry is given by~\cite{Freese:1990rb}
\begin{eqnarray}
 V(\phi)=\Lambda_N^4 \left[1+\cos \left(\frac{\phi }{f_a}\right) \right]\,, 
 \label{eq:axion-potential}
\end{eqnarray}
where $\Lambda_N$ is the $U(1)$ explicit symmetry-breaking energy scale, which determines the inflation scale, 
$f_a$ 
is the axion decay constant, and $\phi$ is the canonically normalized axion field. The first spontaneous $U(1)$ symmetry breaking happens when $T \simeq f_a$ ($T$ represents the temperature of the universe). $f_a$ also reduces the value of the self-coupling of the $\phi$ by $1/f_a$~\cite{Freese:1990rb}. On the other hand, the axionic symmetry protects the flatness of the potential of axionic-inflaton~\cite{Kim:2004rp}, at least up to tree level~\cite{Freese:1990rb}.

The potential of Eq.~\eqref{eq:axion-potential} has a number of discrete maxima at $\phi= n \pi f_a$, with $n=1,3, \cdots$. 
At the vicinity of the location of the maximum of the potential, $\lt|\eta_V\rt|\approx 1/ 2 f_a^2 \ll 1$, and this sets the limit of $f_a$~\cite{Kim:2004rp}. 
This inflationary model, like S-I, comes to an end when $\epsilon_V\sim 1$. 
$n_s$ and $r$ can now be calculated as 
\be
n_s =\frac{f_a^2-2 \, \mpl^2 \, \sec ^2\left(\frac{\phi}{2 f_a}\right)+\mpl^2}{f_a^2}\,, 
\qquad \qquad r= \frac{8 \, \mpl^2 \, \tan ^2\left(\frac{\phi}{2 f_a}\right)}{f_a^2}\,.
\ee 
The values of $n_s$ and $r$ for $10\mpl\lsim f_a \lsim 1585 \mpl$ are shown in Fig.~\ref{Fig:Planck15+BK15-inflation-ns-r-bound} as blue colored region. This inflationary model has been ruled out at $1-\sigma$ CL by \Planck2015+\BICEP2.


\subsection{Hilltop Inflation (H-I)}
\label{Sec:H-I}
%
In this inflationary model (abbreviated as H-I) inflaton starts rolling near the maximum of the potential and this automatically makes $\epsilon_V\sim 0$ at the onset of inflation. The potential is given by~\cite{Kohri:2007gq,Boubekeur:2005zm}
\begin{eqnarray}
 V(\phi)= \Lambda_H^4\left[1-\left(\phi\over v\right)^4 + ...\right]\,,\label{eq:Hilltop-potential}
\end{eqnarray}
where $\Lambda_H$ and 
$v$ are parameters, and the  ellipsis represents the other higher-order terms that make the potential bounded from below, and may be responsible for creating the minimum. The maximum of the potential of Eq.~\eqref{eq:Hilltop-potential} is at $\phi=0$. This inflationary model also ends with $\epsilon_V\sim1$. $n_s$ and $r$ are given by
\ba 
n_s=1-\frac{24 \, \mpl^2 \, \phi ^2 \, \left(v^4+\phi ^4\right)}{\left(v^4-\phi ^4\right)^2}
   \,,
   \qquad r= \frac{128\, \mpl^2\, \phi^6}{\left(v^4-\phi^4\right)^2}\,.
\ea
The values of $n_s$ and $r$ for $0.01\mpl\lsim v \lsim 100\mpl$ exist well inside the $2-\s$ range and are shown as cyan-colored region in Fig.~\ref{Fig:Planck15+BK15-inflation-ns-r-bound}%
\footnote{ For $n_s,r$ predictions of a H-I inflation where potential is bounded from below, see~\cref{Linde's section}.
}.


\subsection{Coleman-Weinberg Inflation (C-I)}
\label{Sec:CI}
This potential of Coleman-Weinberg Inflation (abbreviated as C-I) is actually the effective potential of quartic self-interacting scalar field up to 1-loop order, and it is given by~\cite{Barenboim:2013wra,Okada:2014lxa,Kallosh:2019jnl,Choudhury:2011sq}
\begin{eqnarray}\label{Eq:CW-I-model}
 V(\phi)= \frac{A f^4}{4}+A \phi ^4 \left[\log \left(\frac{\phi }{f}\right)-\frac{1}{4}\right].
\end{eqnarray}
Here $f=\lt<\phi\rt>$ is the renormalization scale and $V(\phi=f)=V'(\phi=f)=0$. $A= A(f)$ is determined by the beta function of the scalar-quartic-coupling with inflaton. Here, we do not go into detailed models of interaction of $\phi$ with other fields, and we take $A$ as a free parameter, and fixed it by the normalization to the amplitude of the \cmb~anisotropies. 
The model predicts $n_s$ and $r$ as
\ba 
n_s &=&1+\frac{32 \, \mpl^2 \, \phi ^2 \left[\left(3 f^4+12 \, \phi ^4 \, \log \left(\frac{f}{\phi }\right)+\phi ^4\right) \log \left(\frac{\phi }{f}\right)+f^4-\phi ^4\right]}{\left(f^4+4
   \phi ^4 \log \left(\frac{\phi }{f}\right)-\phi ^4\right)^2} \,,\\
   r &=& \frac{2048 \,\mpl^2 \,\phi ^6 \,  \log ^2\left(\frac{\phi }{f}\right)}{\left(f^4+4 \,\phi ^4 \, \log \left(\frac{\phi }{f}\right)-\phi ^4\right)^2}\,.
\ea 
In comparison to previous inflationary models, here we are assuming small-filed inflationary scenario with $\phi^*\ll f, \phi_{\rm end}\ll f$ and it ends with $\lt|\eta_V\rt|\sim 1$. 
$n_s-r$ predictions for C-I for different values of $f$ are shown in Fig.~\ref{Fig:Planck15+BK15-inflation-ns-r-bound} as red-colored region%
\footnote{For predictions regarding $f\gsim \mpl$ C-I inflationary model, see~\cref{Linde's section}}.
This model is already ruled even out at $2-\sigma$ level by \Planck2015 data.

Along with the \Planck2015 and \Planck2015+\BICEP2-\KeckArray2015 combined bounds and theoretical predictions for $n_s-r$ for the four above-mentioned inflationary models, Fig.~\ref{Fig:Planck15+BK15-inflation-ns-r-bound} also displays $n_s-r$ contour (brown-colored region on $(n_s,r)$ plane) from Ref.~\cite{Guo:2017qjt} at $1-\sigma$ and $2-\sigma$ CL where $\D\Neff$ is regarded as an independent variable.
To draw this contour, the dataset used are \Planck2015 + \BICEP2 + \KeckArray~B-mode
\cmb~data + baryon acoustic oscillation (BAO) data + direct measured present-day-value of Hubble parameter ${\cal H}_0$ with $\L$CDM$+r+\Neff$ model. They found the best-fit parameter value $\D\neff=0.254$ and $n_s=0.9787$. %
Fig.~\ref{Fig:Planck15+BK15-inflation-ns-r-bound} also exhibits that H-I can satisfy the $(n_s, r)$ values, even in $1-\s$ range for $\ncmb\sim 60$. Predictions from S-I, on the contrary, can fit in $2-\s$ region for larger values of $\ncmb$ while remaining outside of $2-\s$ boundary for $\ncmb\sim 50$. 
\section{Inflaton decay during reheating}
\label{Sec:Inflaton decay during reheating}
Our discussion in the preceding section is independent of the origin of $\D\neff$. %
In this section, we assume that a \bsm~particle which contributes to $\D\neff$, is produced from the inflaton during the reheating epoch%
\footnote{
For BBN and other sources of $\neff$ see Refs.~\cite{Shvartsman:1969mm,doroshkevich1984physical,doroshkevich1984formation,doroshkevich1985fluctuations,doroshkevich1988cosmological,doroshkevich1989large,berezhiani1990physics,berezhiani1991cosmology,sakharov1994horizontal,khlopov2013fundamental,Dolgov:2002wy}, and for a recent review on this, see Ref.~\cite{Sakr:2022ans,Aloni:2023tff}. However, we expect CMB bounds to be the most stringent.
}.
As soon as the slow-roll phase ends, the equation of state parameter becomes $w>-(1/3)$, and inflaton quickly descends to the minimum of the potential and initiates damped coherent oscillations of inflaton about this minimum and the reheating era begins. The energy density of this oscillating field 
is assumed to behave as a non-relativistic fluid with no 
pressure when averaged over a number of coherent oscillations. Therefore, the averaged equation of state parameter during reheating $\bar{w}_r=0$. During this epoch, the energy density of this oscillating inflaton reduces due to Hubble expansion. In addition to that, the energy density of inflaton also decreases owing to interaction with the 
relativistic \sm~Higgs and possibly with a relativistic \bsm~particle, $X$, which eventually contributes to the effective number of neutrinos~\cite{Ichikawa:2007jv}. During the initial phase of this epoch, 
${\cal H}> \G_\phi$. Here, $\G_\phi$ is the 
effective dissipation rate of inflaton, and the small value of $\G_\phi$ is due to the small value of the couplings with inflaton. However, the value of ${\cal H}$ continues to decrease, and soon it becomes ${\cal H}\sim \G_\phi$. Approximately at this time, the energy density of oscillating inflaton and the energy density of relativistic species produced from the decay of inflaton become equal. At this moment, the temperature of the universe, $\Trh$, is given by~\cite{Lozanov:2019jxc,Giudice:2000ex,Garcia:2020wiy}
\eq{\label{Eq:TRH}
\Trh= \lt( \frac{90}{ \gsrh \,\pi^2} \rt)^{1/4} \sqrt{\G_\phi\, \mpl}\,,
}
where $\gsrh$ denotes the effective number of degrees of freedom at the conclusion of the reheating era. We also assume that inflaton decays instantly and completely during the last stage of the reheating era, and the universe thereafter becomes radiation dominated. %
If $\cs_{\rm RH}$ is the cosmological scale factor at the end of reheating, then, the number of $e$-folds during reheating, $\nrh$, is given by~\cite{Drewes:2017fmn}
\ba  \label{Eq:efold-during-reheating}
\nrh \equiv \log\lt(\frac{\cs_{\rm RH}}{\cs_{\rm end}}\rt) =\frac{1}{3} \ln\lt(\frac{\r_{\rm end}}{3 \, \G_\phi^2 \, \mpl^2 } \rt)\,,
\ea  
where $\r_{\rm end}$ is the energy density at the end of inflation. To derive this, we have used Eq.~\eqref{Eq:TRH}, and the fact that the universe becomes radiation dominated at the end of reheating, and also assumed that equation of state does not vary during reheating epoch.
If slow roll inflation ends 
with $\epsilon_{V}(\phi_{\rm end})
=1$ (i.e. with $\dot{\phi}^2|_{\phi=\phi_{\rm end}}=V(\phi_{\rm end})$), then
\begin{align}
\r_{\rm end}= \lt( 1+ \frac{\epsilon_{V}(\phi_{\rm end})}{3} \rt)\, V(\phi_{\rm end})= \frac{4}{3} V(\phi_{\rm end})\,.
\end{align}
Again, rearranging Eq.~\eqref{Eq:efold-during-reheating}, we get
\begin{eqnarray}
\Gamma_\phi=\frac{1}{ \mpl}\left(\frac{\r_{\rm end}}{3}\right)^{1/2} e^{-3\, \nrh/2}\,. \label{Eq:GammaConstraint}
\end{eqnarray}


Now, if 
$\G_{\phi \to XX}$ and $\G_{\phi \to hh}$ are the decay width of inflaton to $X$ and \sm~Higgs particle $h$, then the branching fraction $B_X$ for the production of $X$ particle is defined as
\ba\label{Eq:definition-of-BX}
B_X= \frac{\G_{\phi \to XX}}{\G_{\phi\to XX} + \G_{\phi \to hh}}= \frac{\G_{\phi \to XX}}{\G_\phi} \,.
\ea

Here, $\G_\phi = \G_{\phi\to XX}+ \G_{\phi \to hh}$. 
Now, the time evolution of the energy density of inflaton, $\r_\phi$, energy density of relativistic \sm~particles, $\r_{\rm rad}$, and  energy density of relativistic $X$ particle, $\r_X$, can be computed by solving the following set of differential equations 
\begin{align}
  & \dot\rho_\phi + 3 \,{\cal H} \,\rho_\phi = -\Gamma_\phi \,\rho_\phi\,, 
  \label{Eq:BG1}\\  
  & \dot\rho_{\rm rad} + 4 \, {\cal H} \,\rho_{\rm rad} = \Gamma_\phi\,(1-B_X)\,\rho_\phi\,,  \label{Eq:BG2}\\
  & \dot\rho_X + 4 \,{\cal H}\, \rho_X = \Gamma_\phi \, B_X \,\rho_\phi\,. 
  \label{Eq:BG3}
\end{align}
Here, we assume that $X$ is so feebly interacting with $H$ or other \sm~particles that we can ignore the interaction term. 
Besides, we also assume that $X$-particles are not self-interacting. Therefore, $\r_X$ only decreases due to Hubble-expansion of the universe. 
Furthermore, the Friedmann equation gives
\begin{equation}
  {\cal H}^2 
  =\frac{\rho_\phi + \rho_{\rm rad} + \rho_X}{3 \mpl^2}\,.
  \label{eq_H}
\end{equation}

This $X$ particle fails to establish thermal equilibrium with the thermal \sm~relativistic fluid of the universe, and hence does not share the temperature of photons or neutrinos. This \bsm~article remains relativistic up to today, and thus contributes to the total relativistic energy density of the present universe
\eq{
\r_{{\rm rad},{\rm tot}} &= \r_\g + \r_\n+\r_X\,.       
\label{Eq:Neff-def}
}
Following Sec.~\ref{Sec:Effective number of relativistic degrees of freedom}, we claim that 
$\neff$ takes care of the contribution of $X$ to $\r_{{\rm rad},{\rm tot}}$, 
and in the absence of $X$ particle, 
$\Neff={\neff}_{,\sm}$.
The contribution of $X$ in $\Neff$ is expressed as~\cite{Jinno:2012xb}

\eq{
\D\Neff = \frac{43}{7} \qty[\frac{10.75}{\gssrh 
}]^{1/3} \qty[\frac{\r_X}{\rr}]_{{\cal H}\ll \G_\phi} \,.   \label{Eq:DNeff-for-our-case}
}
The subscript ${\cal H}\ll \G_\phi$ implies that the integration has been done till the reheating process is over. 
In Eq.~\eqref{Eq:DNeff-for-our-case}, $\gssrh$ is the effective degrees of freedom contributing to the entropy density of the universe at the time when $\rho_\phi$ decays completely. We are assuming that $\gssrh\approx 106.75$, and it remains the same throughout the reheating process. %
%
%
%
\begin{figure}[H]
    \centering
    \includegraphics[height=6cm,width=10cm]{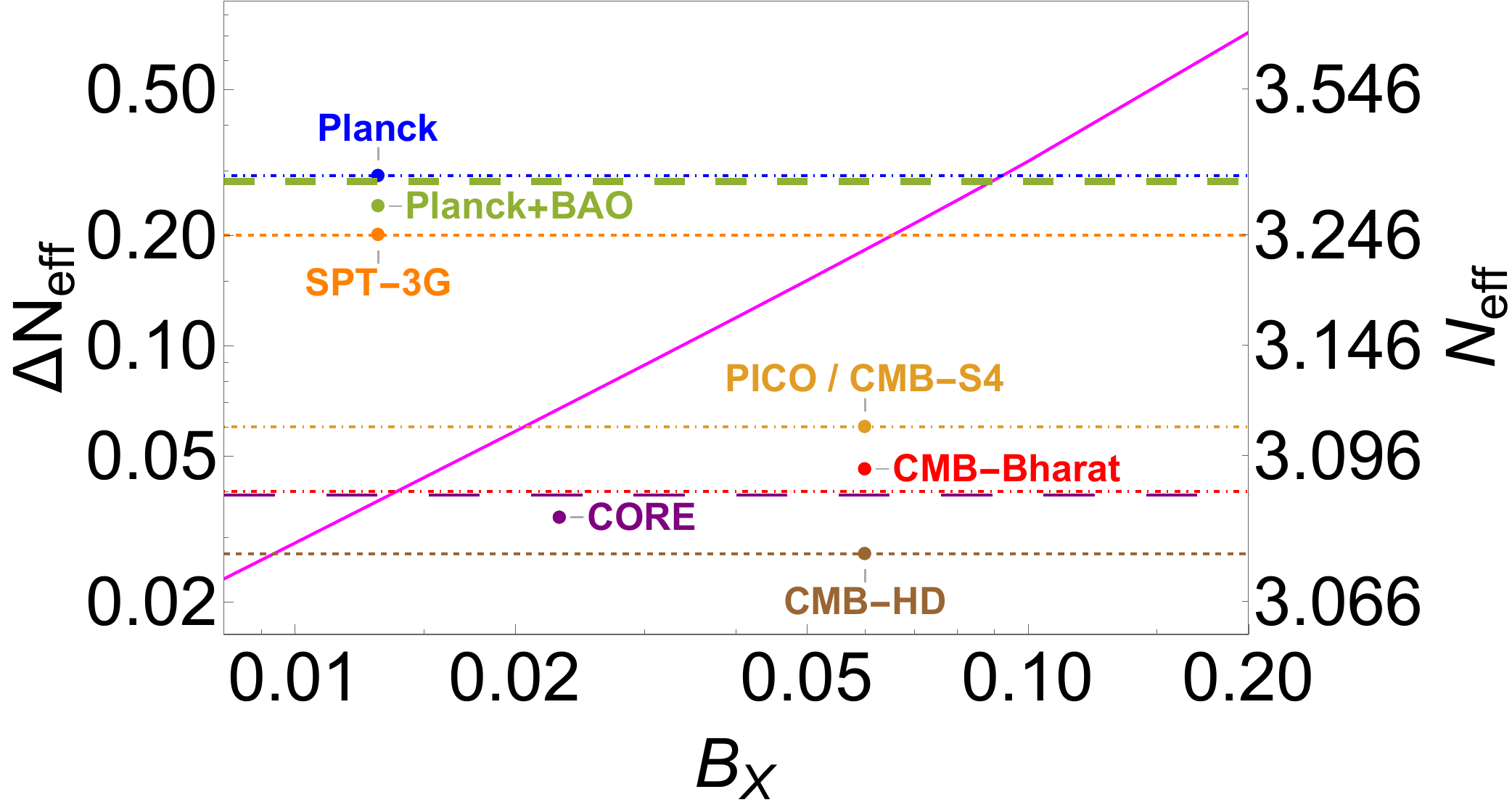}
    \caption{\it \raggedright \label{Fig:Branching-fraction-and-bound-on-DNeff} 
    $\D\Neff$ against $B_X$ following Eq.~\eqref{Eq:DNeff-for-our-case}. The parallel horizontal lines indicate present 
    bounds on $\D\Neff$ and prospective future reaches that will be within the scope of the sensitivity of the upcoming \cmb~observations mentioned in Table.~\ref{Table:bound-on-Delta-Neff}. 
    If the non-thermal $X$-particle is created only from inflaton decay, then it should be $B_X \lsim 0.09$ (\Planck~bound). If 
    $B_X \gsim 0.009$, then it can be tested further by future \cmb~observations.
  }  
\end{figure}%
%
%
%
Now, using the bounds on $\D\Neff$ and prospective future reaches of $\D\neff$ from 
Table~\ref{Table:bound-on-Delta-Neff}, 
$\D\Neff$ as a function of $B_X$ from Eq.~\eqref{Eq:DNeff-for-our-case} (and assuming that $\bx$ does not depend on cosmological scale factor) is shown in Fig.~\ref{Fig:Branching-fraction-and-bound-on-DNeff}. 
Because a higher value of $B_X$ suggests a higher production of $X$ particles, $\D\neff$ grows monotonically with $\bx$. This is true when $\D\Neff$ is solely contributed by $X$ particle.
On the other hand, the terms on the right-hand-side of Eqs.~\eqref{Eq:BG2} and \eqref{Eq:BG3} are the production rate of radiation and $X$, respectively, and only $B_X$ regulates the difference between those two production rates. 
Since $\D\neff\sim \r_X/\r_{\rm rad}$, 
the result shown in Fig.~\ref{Fig:Branching-fraction-and-bound-on-DNeff} is independent of whether $\G_\phi$ is constant or varies with cosmological scale factor and temperature (e.g.~\cite{Barman:2022tzk}). Additionally, this figure depicts that \Planck~and \Planck+BAO bounds draw an upper limit on the possible values of $B_X$. $B_X> 0.09$ is already eliminated by \Planck~data.
%
Furthermore, $0.09\gsim B_X\gsim 0.066$ can be tested by \sptnew, and $B_X\gsim 0.021, B_X\gsim 0.014, B_X\gsim 0.013,$ and $B_X\gsim 0.009$ may be tested from other future \cmb~experiments such as \cmbsfour/PICO, \CMBBharat, \coremfive, and CMB-HD. 

\begin{figure}[H]
    \centering
     \includegraphics[width=0.48\linewidth]{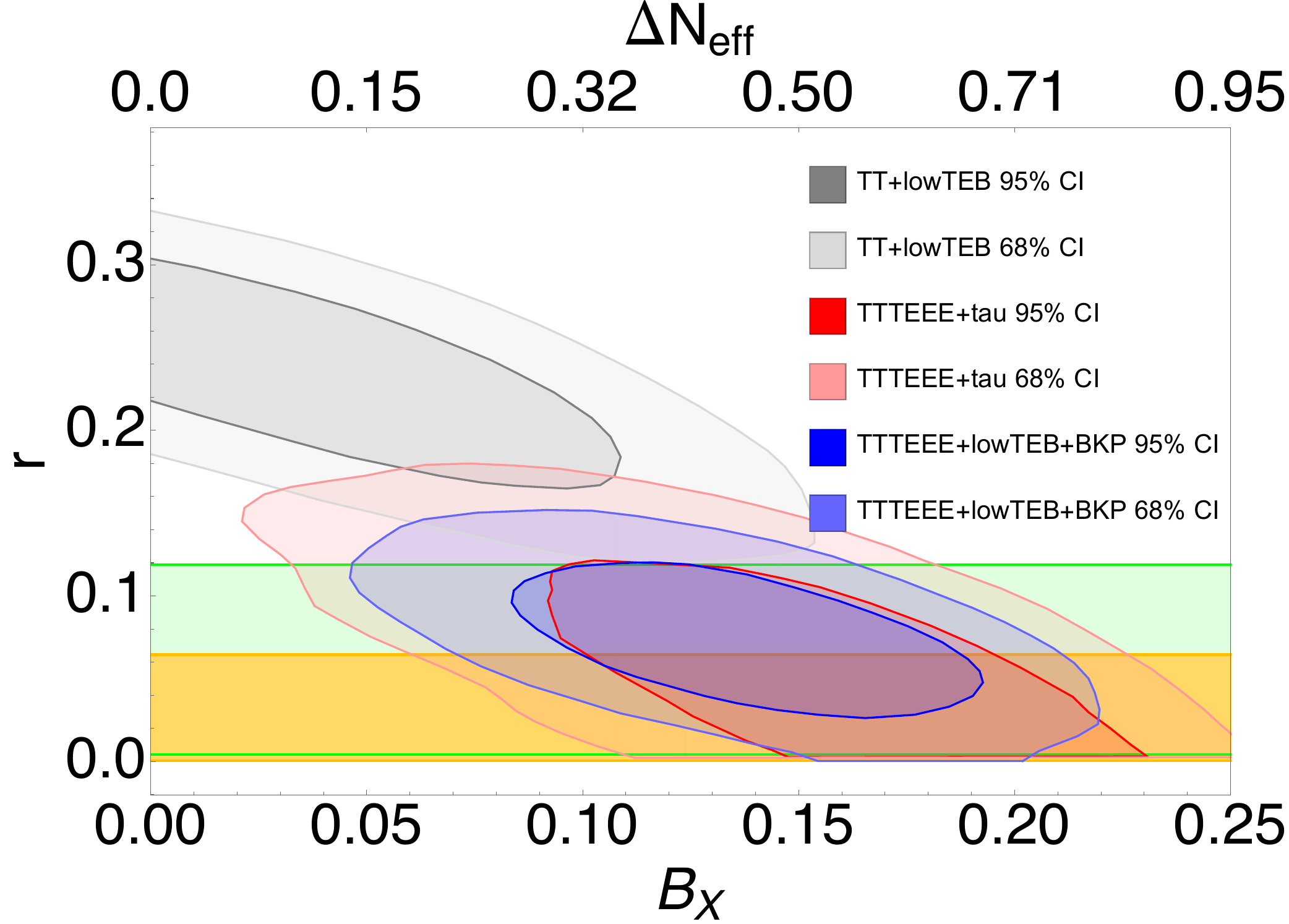}\;
     \includegraphics[width=0.48\linewidth]{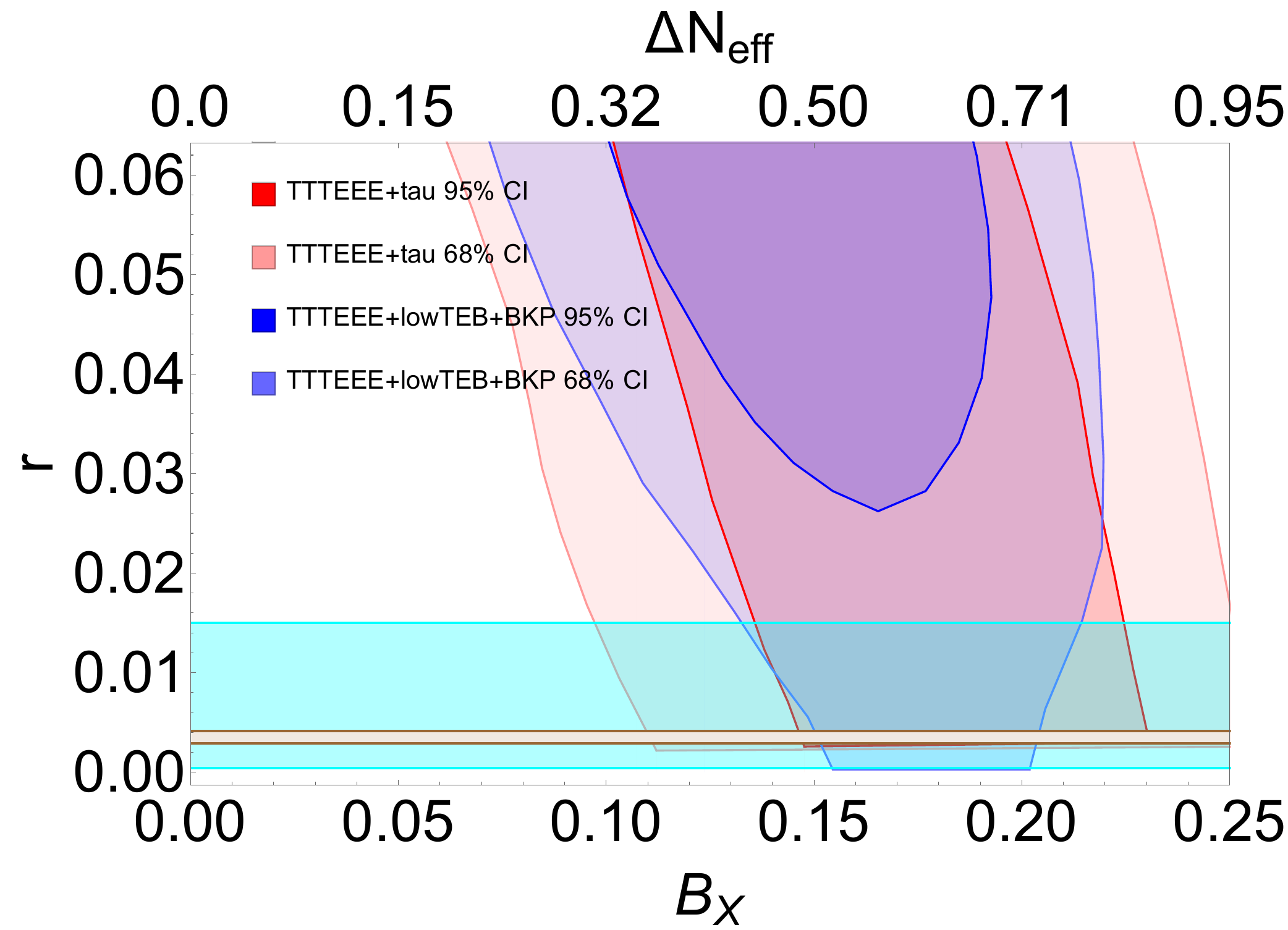}
    \caption{\it \raggedright \label{Fig:r-Br-Bound-region-plot} 
    {\bf Left panel:} Constraints on 
    $(r, B_X)$ plane at $68\%$ and $95\%$ CL from Ref.~\cite{DiValentino:2016ucb} 
    combining \Planck~2015 temperature power spectrum (“PlanckTT + lowTEB”), high multipoles \Planck~polarization data with  \cmb~polarization B modes constraints provided by the 2014 common analysis of \Planck, \BICEP 2 and \KeckArray~(“PlanckTTTEEE + lowTEB+BKP”), and 2016 dataset from \Planck~High Frequency Instrument (HFI) ("tau"). The region covered with light green color indicates the allowed region of $r$ predicted by N-I model, while the yellow-shaded region corresponds to the prediction from H-I inflationary model.
    {\bf Right panel:} same bound on $(r, B_X)$ plane as the left panel, but with predictions of $r$ from C-I (cyan-colored region) and from S-I (brown-colored region). All of these models have been ruled out by `PlanckTT + lowTEB' data on $(r,\bx/\D\neff)$ plane. Although, N-I and H-I remains within $1-\sigma$ limit of both `TTTEEE+tau' and `PlanckTTTEEE + lowTEB+BKP' data, C-I and S-I are within $1-\s$ contour of only `PlanckTTTEEE + lowTEB+BKP' data.
     %
    %
    %
    }  
\end{figure}

We use the relation between $\D\neff$ and $\bx$ from Fig.~\ref{Fig:Branching-fraction-and-bound-on-DNeff} in Fig.~\ref{Fig:r-Br-Bound-region-plot} where allowed ranges of $r$ for the four inflationary models are shown as colored horizontal-stripes on $(r, B_X/\D\Neff)$ plane. On the left panel, the green and yellow colored region exhibits allowed range for $r$ for N-I and H-I models. The cyan and brown colored region on the right panel of Fig.~\ref{Fig:r-Br-Bound-region-plot} illustrates the allowed range for $r$ for the other two inflationary models – C-I and S-I, respectively. The $68\%$ and $95\%$ CL contours depicted on the background on this $(r, B_X/\D\Neff)$ plane are taken from~\Ccite{DiValentino:2016ucb}. These 2-dimensional contours are from $8$-parameter analysis, including $r$ and $\Neff$, of \Planck~and \BICEP~data, assuming $\Lambda$CDM and flat universe.  
The datasets used here are the combined \Planck2015 temperature power spectrum ($2<\ell<2500$, $\ell$ is the multipole number, and indicates an angular scale on the sky of roughly $\pi/\ell$~\cite{Bambi:2015mba}) with  polarization power spectra for $(2<\ell<29)$ (“PlanckTT + lowTEB”)~\cite{Planck:2015bpv}, high multipoles \Planck~polarization data with \cmb~polarization B modes constraints provided by the 2014 common analysis of \Planck2015, \BICEP2 and \KeckArray~(“PlanckTTTEEE + lowTEB+BKP”)~\cite{BICEP2:2015nss}, and 2016 dataset from \Planck~High Frequency Instrument (HFI) ("tau")~\cite{Planck:2016kqe}. The best fit value of $\D\neff$ at $68\%$ CL are $0.504$ for “PlanckTT + lowTEB”, $0.094$ for “TTTEEE+tau”, $0.464 $ for “PlanckTTTEEE + lowTEB+BKP”.~%
Regarding PlanckTT + lowTEB dataset, all of these four inflationary models are ruled out at more than $2-\s$.  
%
%
%
%
 Fig.~\ref{Fig:r-Br-Bound-region-plot} also shows that four inflationary models will be within $2-\s$ contour of `TTTEEE+tau' data if $\bx \gsim 0.033$ (for N-I), $\bx\gsim 0.061$ (for H-I), $\bx\gsim 0.097$ (for C-I), $\bx\gsim 0.11$ (for S-I). Similarly, to be within $2-\s$ contour of `TTTEEE+lowTEB+BKP' it is required that $\bx\gsim 0.046$ (for N-I), $\bx\gsim0.07$  (for H-I), $\bx\gsim 0.13$  (for C-I), $\bx\gsim 0.15$ (for S-I). 
By using this now along with the Fig.~\ref{Fig:Branching-fraction-and-bound-on-DNeff}, it is possible to determine whether any inflationary model is compatible with the  assumption that inflaton is the only source of $X$ and this $X$ contributes solely and entirely to $\neff$. For instance, S-I and C-I inflation are incompatible with the aforementioned assumption 
regarding the `TTTEEE+tau' and `TTTEEE+lowTEB+BKP' dataset. If $X$ originates from a different source, these inflationary models can still remain inside $1-\s$ contour of `TTTEEE+tau' data. In contrary, H-I is compatible with the assumption that inflaton is the sole source of $X$, at least, up to $2-\s$ CL interval. 
 It is also worth noting that, the best fit value of $n_s$ varies in presence of $\neff$, just like the best-fit value of $r$. As a result, Fig.~\ref{Fig:r-Br-Bound-region-plot} cannot be utilized to make the final decision to validate any inflationary models. In Fig.~\ref{Fig:r-Br-Bound-region-plot}, the N-I model, for example, predicts r value within $2-\s$ bounds, but this model is already ruled out in Fig.~\ref{Fig:Planck15+BK15-inflation-ns-r-bound} due to the predicted small value of $n_s$.



\section{Inflaton Decaying to Dark Radiation}
\label{Sec:Inflaton Decay}
Following the preceding section, we postulate that during reheating, the inflaton decays to \sm~Higgs doublet $H$ and \bsm~particle $X$. 
The Lagrangian density describing such a process can be expressed as~\cite{Drewes:2017fmn,Drewes:2019rxn,Lozanov:2019jxc,Allahverdi:2010xz}
\be\label{Eq:BSM+Higgs-production-lagrangian}
{\cal L}\supset -  \lambda_{12, H} \, \s_m \, \phi \, H^\dagger H  - {\cal L}_{\phi\to X} 
\, - {\cal L}_{\rm scattering}
\,, 
\ee  
where 
the first term on the right side of Eq.~\eqref{Eq:BSM+Higgs-production-lagrangian} 
encodes the decay of $\phi$ to the \sm~Higgs particle $h$, and hence this term is accountable for the generation of thermal relativistic \sm~plasma of the universe. $\lambda_{12, H}$ is dimensionless in this case, and $\s_m$, the mass scale, is taken to be equal to $\mphi$.
Subsequently, the decay width to \sm~Higgs particle $h$ is
\begin{align}\label{Eq:Inflaton decaying to two higgs}
\Gamma_{\phi\to {h h}}  \simeq \frac{\lambda_{12,H}^2 \, \s_m^2}{8\pi\, \mphi} 
=\frac{\lambda_{12,H}^2 \, \mphi}{8\pi\,}
\,,
\end{align}
%

%
In~\cref{Sec:SI,Sec:NI,Sec:H-I,Sec:CI}, we have discussed four inflationary scenarios. 
The minimum of the potential of N-I is located at $f\pi$, whereas for S-I, it is at $\phi=0$, and for C-I, it is at $\phi = f$.
Then, the masses of inflaton for the aforementioned three inflationary models are
\begin{align}
    m_\phi=
    \begin{cases}
      \frac{2}{\sqrt{3}}\frac{\Lambda^2}{M_P} \qquad &\text{(for S-I~\cite{DiValentino:2016nni})}\,,\\
        \frac{\Lambda_N^2}{f_a} \, \qquad &\text{(for N-I~\cite{Freese:2014nla})} \,,\\
        2\sqrt{A} f \qquad &\text{(for C-I~\cite{Panotopoulos:2021ttt})}\,.
    \end{cases}
\end{align}
Since, $\Lambda, \L_N, f_a, A$, and $f$ are determined from the best-fit value of $n_s,r, A_s$ obtained from CMB data and the number of $e$-folds ${\cal N}_{\rm CMB}$, $m_\phi$ for the aforementioned three inflationary models are determined by the data from CMB observations. However, it should be noted that N-I and C-I are already in discordance with $n_s-r$ contour from BICEP data, even at $2-\sigma$ CL (see~\cref{Fig:Planck15+BK15-inflation-ns-r-bound,Fig:[Newest]inflation-ns-r-bound}). Furthermore, the form of the potential of Hilltop model, as described in~\cref{eq:Hilltop-potential}, is expected to be bounded from below by other terms without affecting the inflationary predictions. Therefore, it becomes necessary to consider a specific theory to define the mass of the inflaton in this scenario (See Ref.~\cite{Kallosh:2019jnl}, for further details).

To keep the discussion of this section more general, we consider a broader range of inflationary scenarios beyond the four inflationary scenarios discussed above. For example, if we consider quartic inflationary potential, we need to add a bare mass term to study reheating (for example, see.~\Ccite{Dimopoulos:2017xox}). This is why, instead of fixing a specific value, we explore variations in the $m_\phi$ during our analysis.

Furthermore, ${\cal L}_{\phi\to X}$ in Eq.~\eqref{Eq:BSM+Higgs-production-lagrangian} is the interaction term of $\phi$ with $X$. Let us also assume that $\lambda_{\phi X}$ is the coupling of inflaton-\bsm~particle interaction. Now, to make the discussion as generic as feasible, we suppose that $X$ can be a light fermion $\chi$, a \bsm~scalar $\varphi$, or a $U(1)$ gauge boson $\ug$. Therefore, possible interaction Lagrangian with $\phi$ includes~\cite{Drewes:2019rxn,Drewes:2017fmn}

\begin{empheq}[
  left=
    {{\cal L}_{\phi\to X} \supset } 
    \empheqlbrace
]{align}
&+y_{\chi} \, \phi \, \bar{\chi}\chi \,, &&\lt(\phi \to  \bar{\chi}\chi\,; \quad 
\lpx\equiv y_\chi
\rt) \,, \label{Eq:DecayToFermion}\\
&+ \lambda_{12,\varphi} \, \s^\prime_m 
\,\phi \, \varphi\varphi \,, &&\lt(\phi \to \varphi\varphi  
\,; \quad \lpx\equiv  \lambda_{12,\varphi}
\rt)\,,\label{Eq:DecayToBoson}\\
  %
  %
&+ \frac{\lambda_{13,\varphi}}{3!} \, \phi \, \varphi\varphi\varphi \,,  &&\lt(\phi \to \varphi\varphi\varphi  
\,; \quad \lpx\equiv \lambda_{13,\varphi}
\rt)\,,\label{Eq:QuibicInteractionToBoson}\\
&+ \frac{g_{\phi \ug}}{\L_m} \, \phi \, F_{\m\n}\tilde{F}^{\m\n} \,, &&\lt(\phi \to \ug\ug  
\,; \quad \lpx\equiv g_{\phi \ug}
\rt) \,. \label{Eq:DecayToGaugeBoson}
\end{empheq}
Here $y_\chi, \,\lambda_{12,\varphi}, \, 
\lambda_{13,\varphi},$ and $ g_{\phi\ug} $ are dimensionless couplings with $\s^\prime_m, \L_m$ as mass scales. $F_{\m\n}$ is the field strength tensor of $\ug$, and $\tilde{F}^{\m\n} $ is its dual. 
Additionally, ${\cal L}_{\rm scattering}$ in~\cref{Eq:BSM+Higgs-production-lagrangian} represents the scattering terms involving inflaton with both \sm~and \bsm~particles, and can be written as follows
\begin{align}\label{Eq:new scat lag}
    {\cal L}_{\rm scattering}&= \lambda_{22} \, \phi\phi \, H^\dagger H + {\cal L}_{\phi\phi\to X X} +\text{h.c.}\,,
\end{align}
where $\lambda_{22}$ is dimensionless couplings. We assume that ${\cal L}_{\phi\phi\to X X}$-channel is not effective in contribution to $\neff$. For instance, for $\phi \phi \to \varphi\varphi $ channel, we can write~\cite{Drewes:2019rxn}
\begin{align}
{\cal L}_{\phi\phi\to X X} =+ \frac{\lambda_{22,\varphi}}{4} \, \phi\phi \, \varphi\varphi \,, \qquad &\lt(\phi \phi \to \varphi\varphi  
\,; \quad \lpx\equiv \lambda_{22,\varphi}  \rt)   \label{Eq:ScatteringToBoson}\,.
\end{align}
In~\cref{Sec:Boltzman solve for scalar DR scattering}, we show that the contribution of $\varphi$ producing via~\cref{Eq:ScatteringToBoson} in $\D\neff$ is negligible compared to $\varphi$ produced via $\phi\to \varphi\varphi$ decay channel, due to the reaction rate in the former case being dependent on the instantaneous value of inflaton energy density. 
In~\cref{Sec:thermal equilibrium fish}, we explore the range of these couplings to investigate whether \dr~can achieve thermal equilibrium with \sm~Higgs via inflaton exchange processes.

\begin{figure}[htp!]
    \centering    
\includegraphics[height=6cm,width=8cm]
{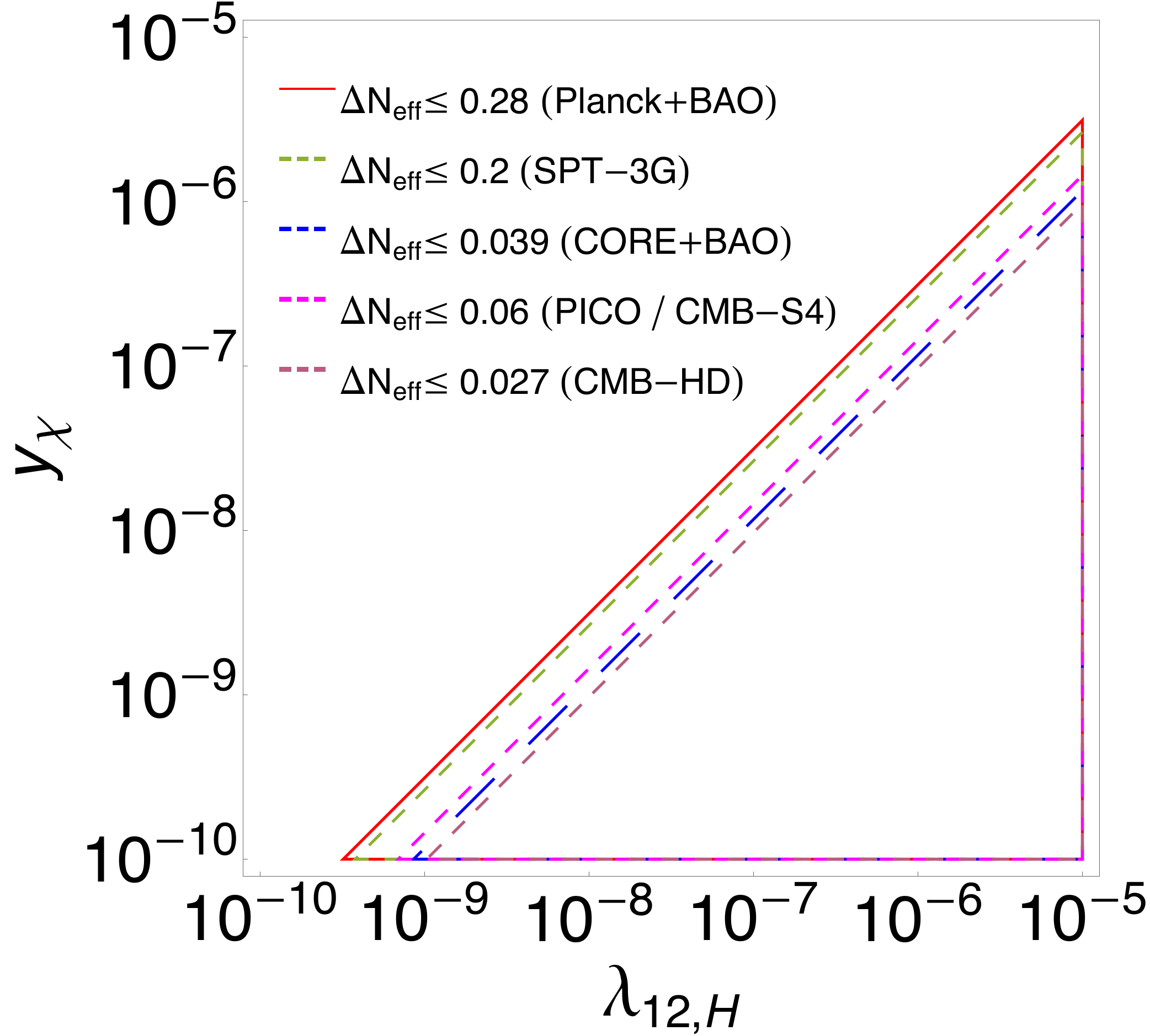}\;
%
\includegraphics[height=6cm,width=8cm]
{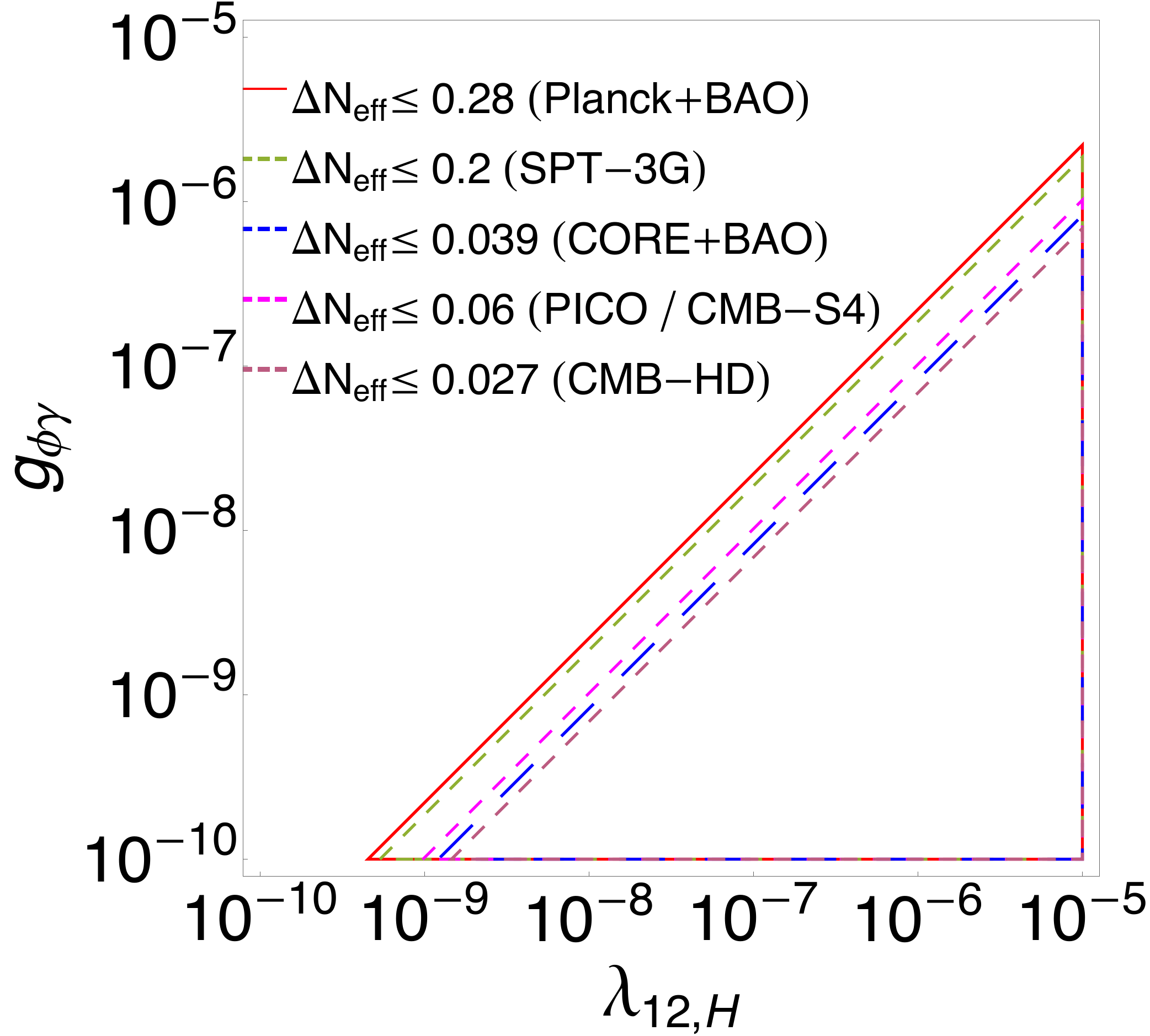}
\includegraphics[height=6cm,width=8cm]
{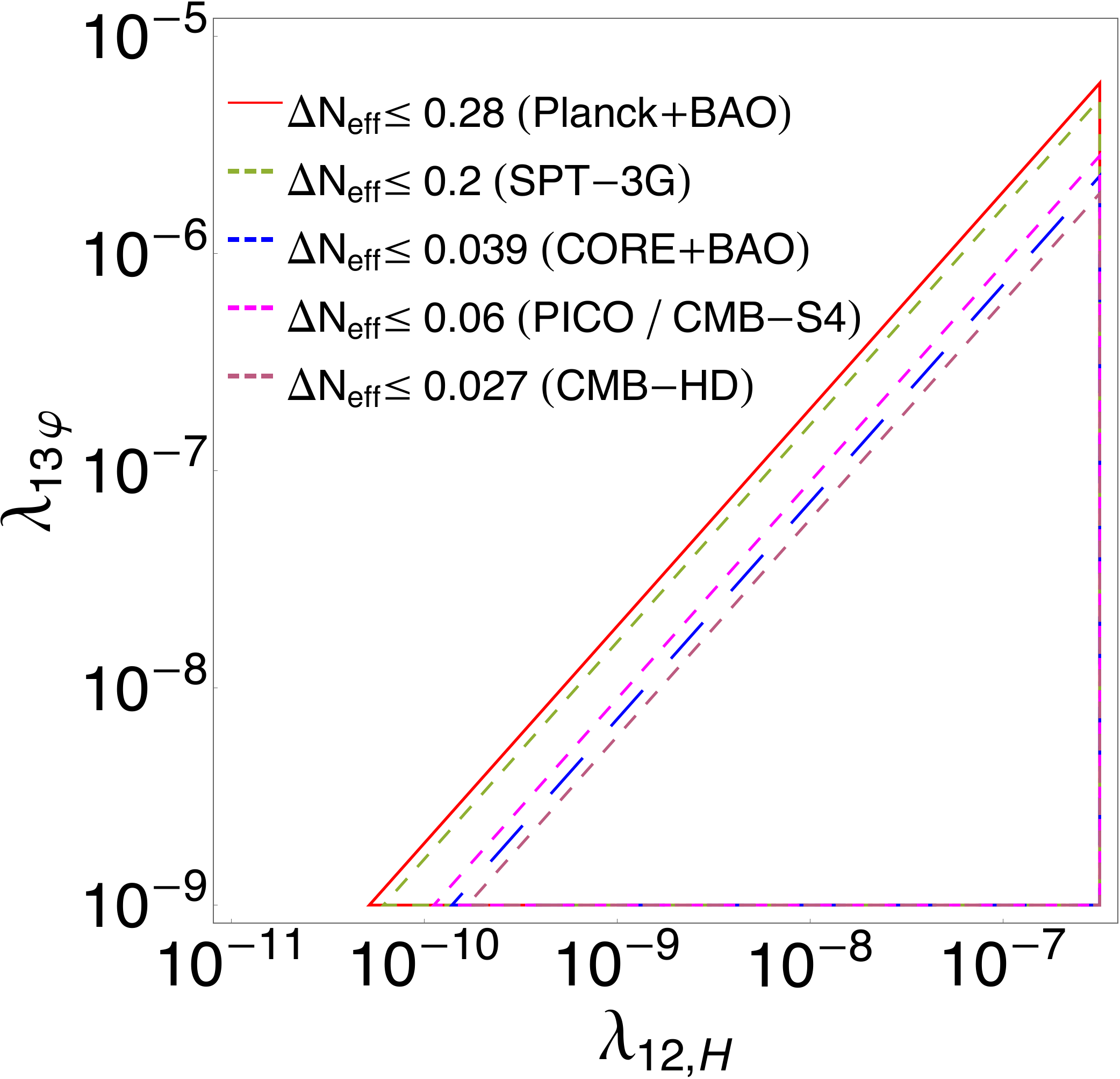}
    \caption{\it \raggedright \label{Fig:Allowed-region-Plane-of-TwoCouplings} %
    Colored lines delimit regions 
    corresponding to the value of $\D\Neff \le$ current bounds and prospective future reaches that will be within the scope of upcoming \cmb~experiments, mentioned in Table.~\ref{Table:bound-on-Delta-Neff}. The solid line is for \Planck+BAO bound, and the region above this line is already ruled out. The dashed-lines are for prospective future reaches that will be within the scope of exploration from upcoming \cmb~observations and the parameter space above those dashed-lines could potentially be measured by upcoming \cmb~observations. Different colors have been used to represent different bounds from different \cmb~observations. \textbf{Top-left panel} displays such regions for $\phi\to \tilde{\chi}\chi$ on $(y_\chi,\lambda_{12,H})$ plane. Similar regions can be obtained for $(\phi\to \varphi\varphi)$ on $(\lambda_{12,\varphi},\lambda_{12,H})$ plane for $\s^\prime_m=\mphi$ (see the similarity between the branching fraction of Eq.~\eqref{Eq:BX-Decay-To-Fermion} and Eq.~\eqref{Eq:BX-Decay-To-Boson}).
    \textbf{Top-right panel} and \textbf{bottom panel} illustrate such regions for $\phi\to \ug\ug$ on $(g_{\phi\ug}, \lambda_{12,H})$ plane and for $\phi\to \varphi\varphi\varphi$ on $(\lambda_{13,\varphi}, \lambda_{12,H})$ plane, respectively. 
    }
\end{figure}

Now, the reaction rates for the interactions from~\cref{Eq:DecayToBoson,Eq:DecayToFermion,Eq:DecayToGaugeBoson}, in leading order, and related branching fractions are as follows: 
\eq{
&\G_{\phi\to \chi\chi} \simeq \frac{y_\chi^2\, m_{\phi }}{8\pi}\,, && B_\chi = \frac{  \lt(y_\chi/\lambda_{12,H} \rt)^2}{ 1+ \lt( y_\chi/ \lambda_{12,H} \rt)^2  }\, 
&\lt(\text{If}\,X\equiv\chi,\, \text{and}\, \phi \to  \bar{\chi}\chi \rt),       \label{Eq:BX-Decay-To-Fermion}
\\
&\G_{\phi\to\varphi\varphi} \simeq \frac{\lambda_{12,\varphi}^2\, {\s^\prime_m}^2}{8\pi\,m_{\phi}}\,, 
&& B_\varphi = \frac{  \lt(\lambda_{12,\varphi}/\lambda_{12,H} \rt)^2}{ \lt(\mphi/{\s^\prime_m}\rt)^2+ \lt( \lambda_{12,\varphi}/ \lambda_{12,H} \rt)^2  }\, &\lt(\text{If}\, X\equiv\varphi,\,\text{and}\,   \phi \to \varphi\varphi\rt),  \label{Eq:BX-Decay-To-Boson}\\
%
%
&\G_{\phi\to \varphi\varphi\varphi}=\frac{\lambda_{13,\varphi}^2 \, \mphi}{3!64 (2\pi)^3} \,, 
&& B_\varphi=\frac{  \lt(\lambda_{13,\varphi}/\lambda_{12,H} \rt)^2}{ 384\pi^2+ \lt( \lambda_{13,\varphi}/ \lambda_{12,H} \rt)^2  }\,
&\lt(\text{If}\, X\equiv\varphi,\,\text{and}\,   \phi \to \varphi\varphi\varphi\rt),\label{Eq:BX-Cubic-Boson}\\
&\G_{\phi\to \ug\ug}=\frac{g_{\phi \ug}^2}{4 \pi\, \L_m^2} \, \mphi^3 \,, 
&& \hspace{-0.2cm} B_\ug = \frac{  \lt(g_{\phi\ug}/\lambda_{12,H} \rt)^2}{  (1/2)\lt(\L_m/\mphi\rt)^2\, +  \lt( g_{\phi\ug}/ \lambda_{12,H} \rt)^2  }\,
\hspace{-0.7cm}
&\lt(\text{If}\,X\equiv\ug, \,\text{and}\,   \phi \to \ug \ug\rt).   \label{Eq:BX-Decay-To-Gauge-Boson}
}
In this work, we assume $\s^\prime_m=\epsilon_1\, \mphi$ and $\L_m=\epsilon_2\, \mphi$ where $\epsilon_1,\epsilon_2$ are dimensionless and $\subset \mathbb{R}_{> 0}$. %
Consequently, $B_\chi$, $B_\varphi$ (from Eqs.~\eqref{Eq:BX-Decay-To-Boson} and~\eqref{Eq:BX-Cubic-Boson}), and $B_\ug$ are independent of $\mphi$. 
Then, in Fig.~\ref{Fig:Allowed-region-Plane-of-TwoCouplings},
we explore the allowable area on the 2-dimensional parameter-space of $(\lpx, \lh)$ 
for the interactions from Eqs.~\eqref{Eq:DecayToFermion}-\eqref{Eq:DecayToGaugeBoson},
regarding the present bounds on $\D\neff$ and prospective future reaches of $\D\neff$ that will be within the scope of future exploration by upcoming \cmb~observations listed in Table~\ref{Table:bound-on-Delta-Neff}. Here, we set all the mass scales to $\s^\prime_m=\L_m=\mphi$. Furthermore, in order to prevent nonperturbative particle production from becoming significant during reheating, we maintain the values of the couplings below ${\cal O}(10^{-4})$~\cite{Ghoshal:2022fud,Drewes:2019rxn}.
We observe that for a given value of $\D\neff$, a greater value of $\lh$ implies a higher value of the coupling of the inflaton with the \bsm~particle. 
This is to be expected because a higher value of $\lh$ implies more generation of \sm~relativistic particles. To maintain $\neff$ constant, a greater value of the coupling of the dark sector to inflaton is required.
The red solid line stands for $\D\neff=0.28$ implying that \Planck+BAO rules out the region above this line. The dashed lines are prospective future reaches that will be within the range of exploration from upcoming \cmb~experiments with higher sensitivity. We do not present the region plot separately for $\phi\to\varphi\varphi$ since $B_\varphi$ for this process on $(\lambda_{12,\varphi},\lh)$ plane has similar form of $B_\chi$ on $(y_\chi,\lh)$ plane. However, following the discussion of~\cref{Sec:thermal equilibrium fish}, we should consider a lower range of $\lambda_{12, \varphi}$ to prevent the scalar \dr~particles  from reaching thermal equilibrium with \sm~Higgs. 
To remain within \Planck~bound, $y_\chi/\lh\lsim 0.31$.  Contrarily, for $\phi\to\varphi\varphi\varphi$, $\lambda_{13,\varphi}/\lh \lsim 19.37$ and for $\phi\to\ug\ug$, $g_{\phi\ug}/\lh \lsim 0.23$. The presence of $1/2$ and $384\pi^2$ in the denominator of the branching fraction in Eqs.~\eqref{Eq:BX-Cubic-Boson} and~\eqref{Eq:BX-Decay-To-Gauge-Boson} results in these differing upper bounds on the coupling ratios. 

%
%


%
\begin{figure}[htp!]
    \centering    
\includegraphics[height=6cm,width=8cm]
{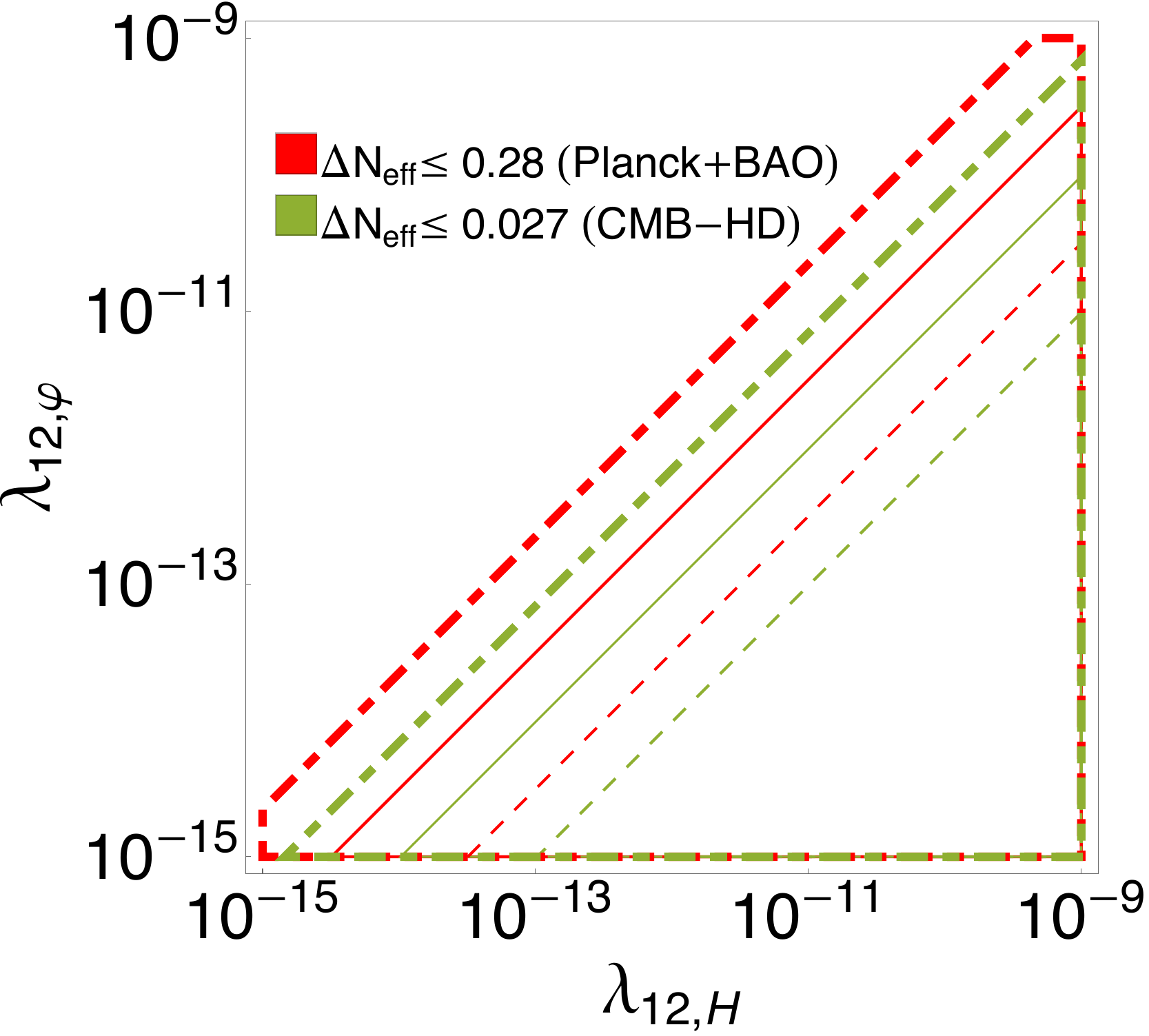}\;
%
\includegraphics[height=6cm,width=8cm]
{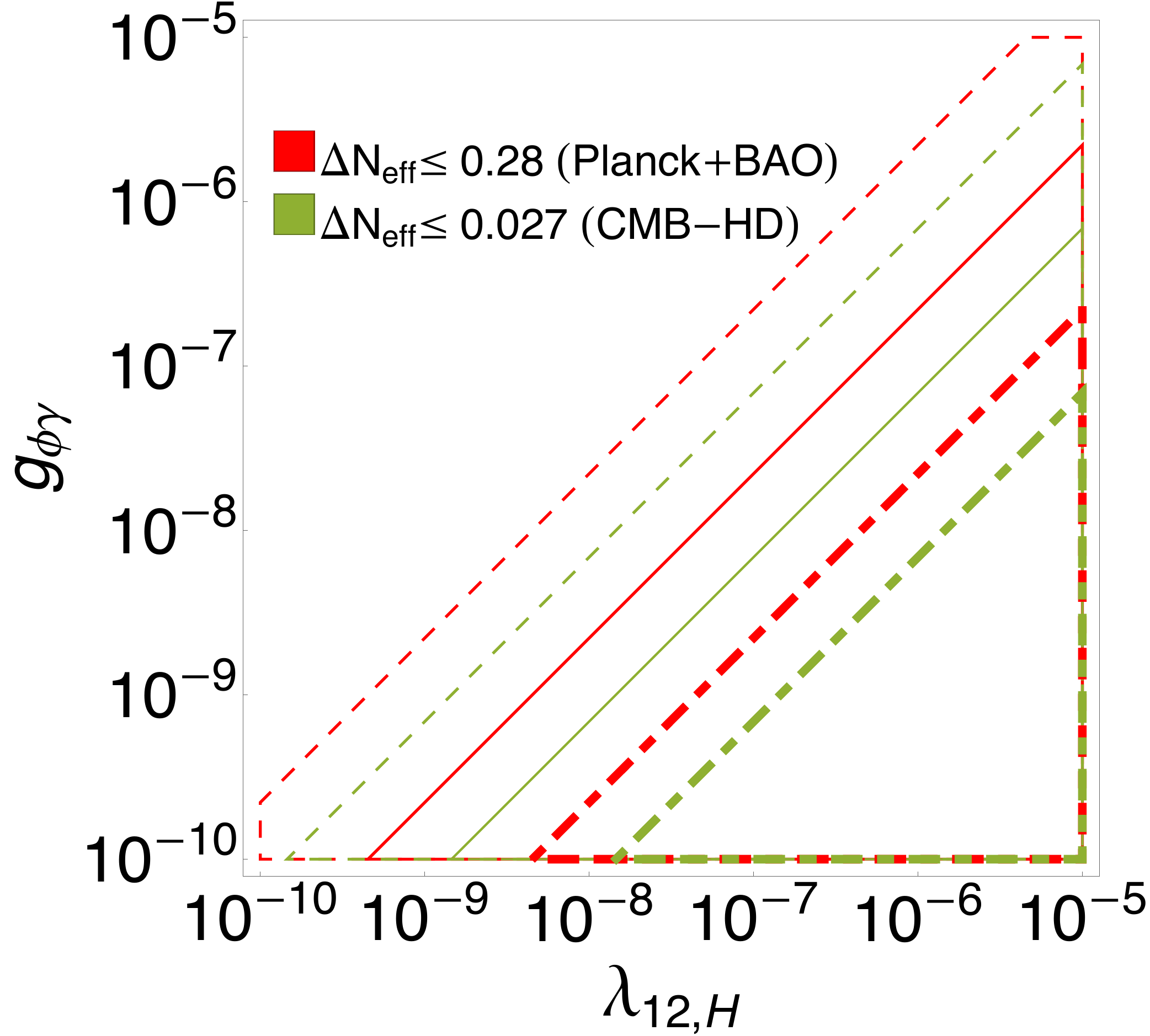}
    \caption{\it \raggedright \label{Fig:mass-scale-varied} %
    Illustration of the alteration of the permissible area on $(\lpx, \lh)$ plane when $\s^\prime_m,\L_m \neq \mphi$. Solid lines represent $\s^\prime_m=\L_m=\mphi$ whereas dashed and dashed-dot lines indicate $\s^\prime_m=\L_m=10\, \mphi$ and $\s^\prime_m=\L_m=0.1 \, \mphi$, respectively. \tb{Left-panel} is for $\phi\to \vp\vp$ and \tb{right panel} is for $\phi\to \ug\ug$. Red color lines are corresponded to $\D\neff=0.28$ (current bound from \Planck+BAO) and green colored lines belong to $\D\neff=0.027$ (prospective future reach of $\D\neff$ of upcoming observation – CMB-HD). The area above the lines are excluded (within the  future reach) by \Planck~(CMB-HD).
   }
\end{figure}
%

In Fig.~\ref{Fig:mass-scale-varied}, we consider current bound on $\D\neff$ - bounds from \Planck~and only one prospective future reach of $\D\neff$ that could be observed by CMB-HD, to demonstrate the shift in the allowable region on the respective $(\lpx, \lh)$ space for the choice of respective mass scales either greater or smaller than $\mphi$ for $\phi\to \varphi\varphi$ and $\phi\to \ug\ug$. Similar to~\cref{Fig:Allowed-region-Plane-of-TwoCouplings}, here also the area above the lines are excluded or to be identified by present or future \cmb~observations. 
If we choose $\s^\prime_m>\mphi$ and $\s^\prime_m<\mphi$, the allowed area decreases and increases for $\phi\to \vp\vp$ and it is displayed on the left panel of Fig.~\ref{Fig:mass-scale-varied}. Contrarily, the right panel of this figure exhibits that increment or reduction of the permissible area for the choice of $\L_m$ less or larger than $\mphi$. This conclusion is in congruence with the expression of $B_\vp$ and $B_\ug$ from Eqs.~\eqref{Eq:BX-Decay-To-Boson} and~\eqref{Eq:BX-Decay-To-Gauge-Boson}.

\begin{figure}[H]
    \centering    
\includegraphics[height=6cm,width=8cm]{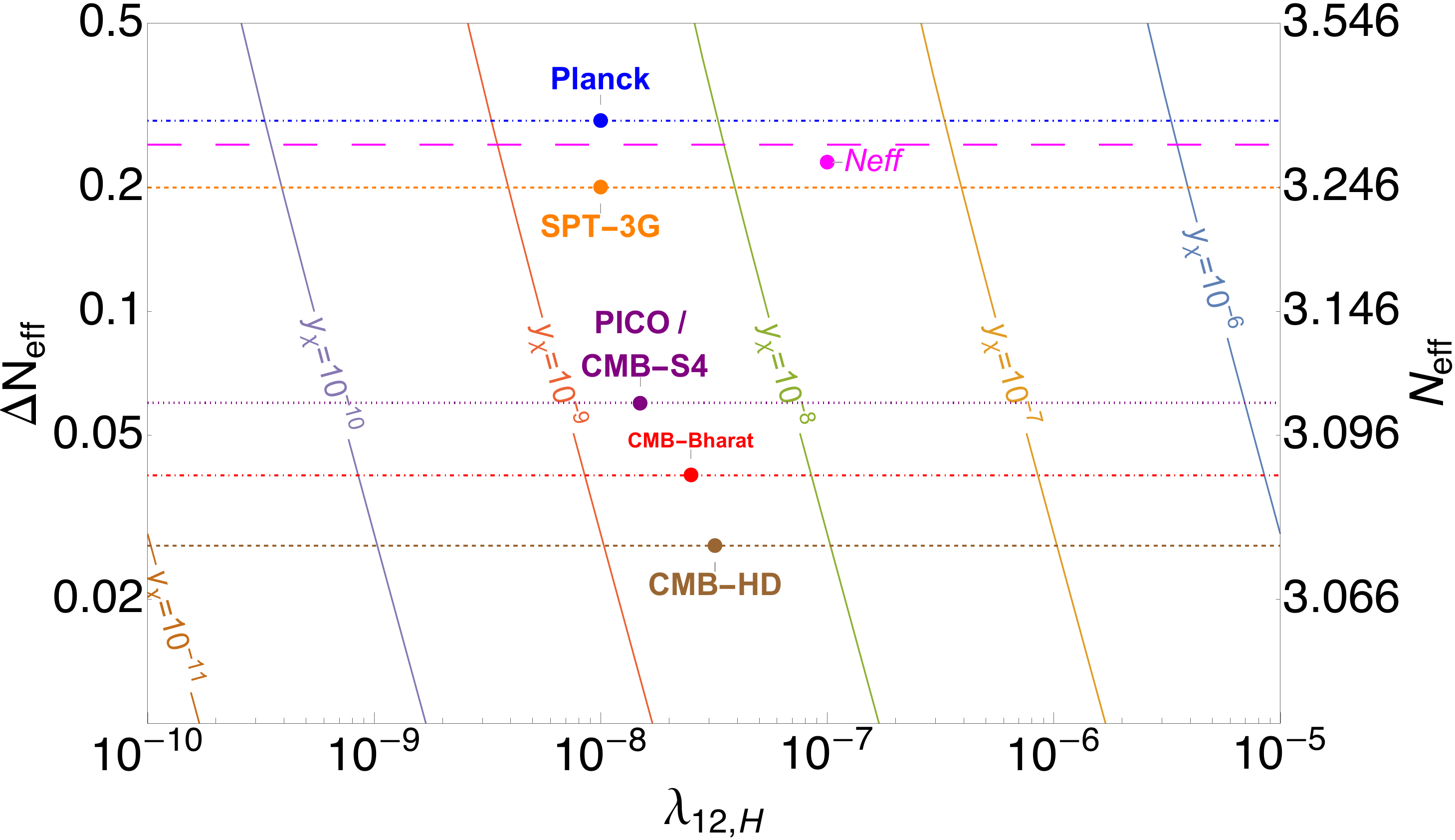}\;
\includegraphics[height=6cm,width=8cm]{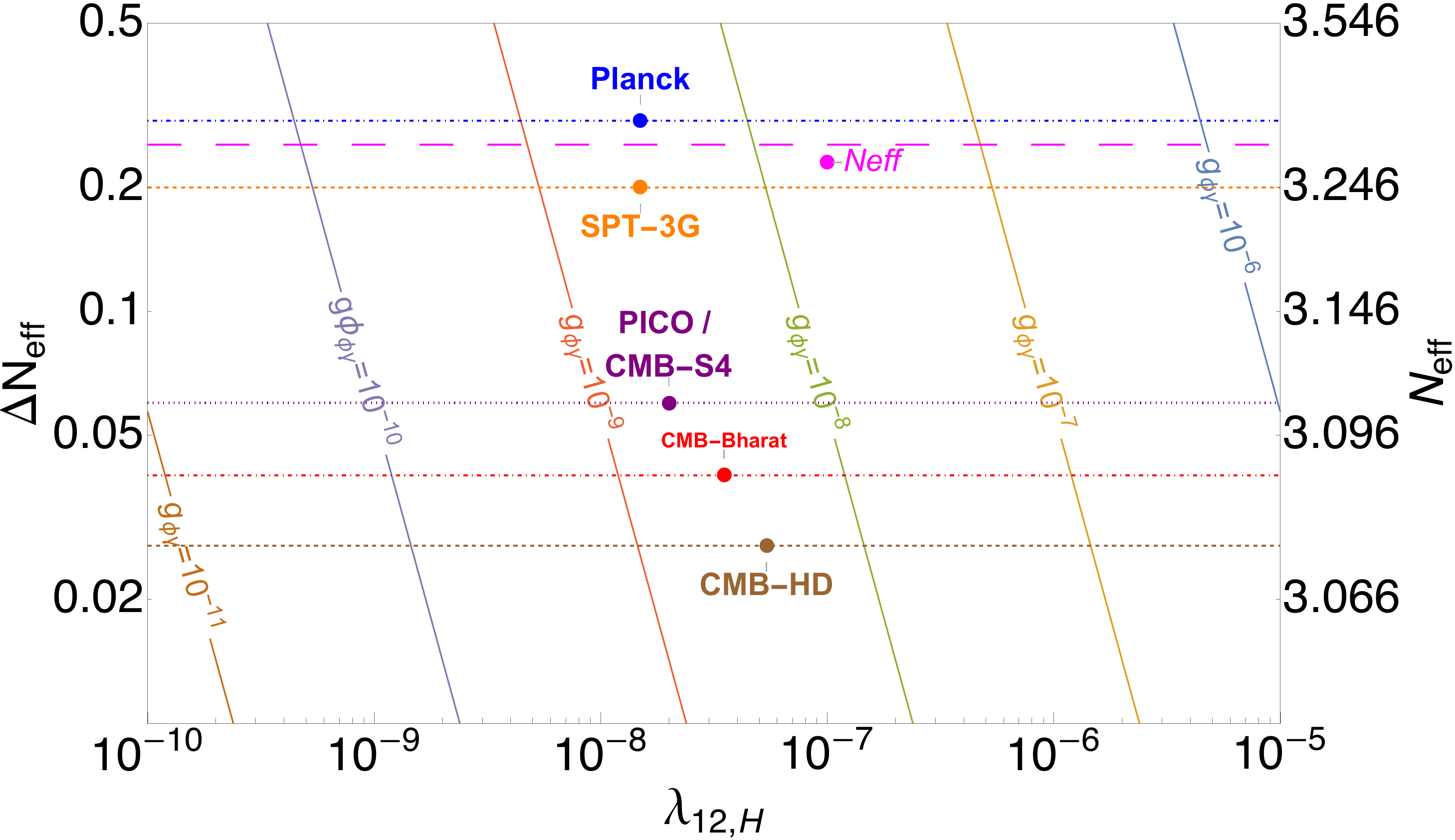}
\includegraphics[height=6cm,width=8cm]{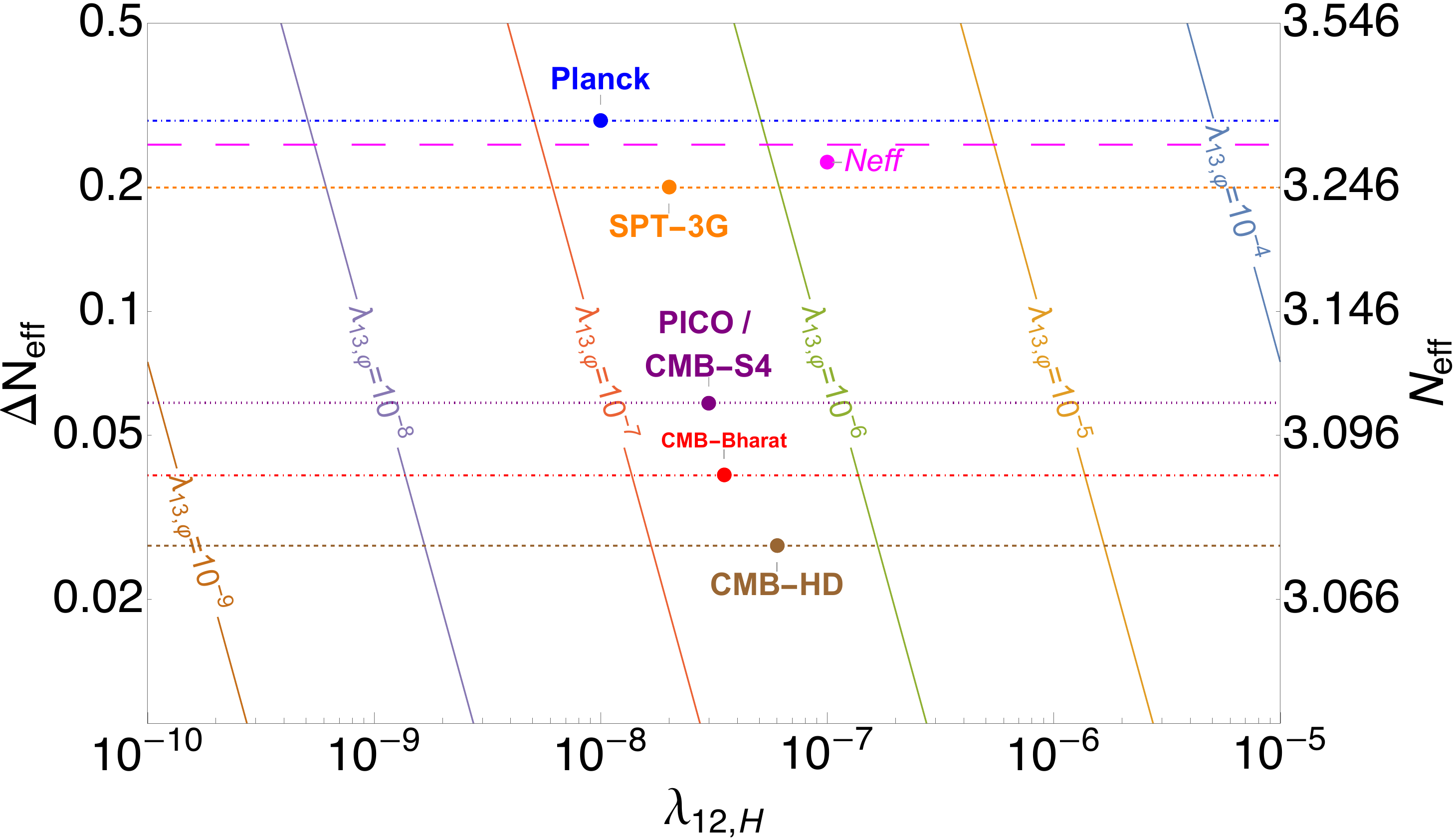}
    \caption{\it \raggedright \label{Fig:DNeff-vs-L12}
      Inclined lines correspond to fixed values of the inflaton-\bsm~particle couplings on $(\D\Neff, \lh)$ plane. For this figure, we set all mass scales $\sim \mphi$. 
      The \cmb~experimental reaches on $\D\Neff$ from \cref{Table:bound-on-Delta-Neff} are shown by discontinuous-horizontal lines. The dashed line in magenta color indicates $\D\neff=0.254$, the value predicted by Ref.~\cite{Guo:2017qjt}. 
      \tb{Top-left panel}, \tb{top-right panel} and \tb{bottom panel} are for $\phi\to \tilde{\chi}\chi$, $\phi\to\ug\ug$, and for $\phi\to\vp \vp\vp$, respectively.
      }
\end{figure}

Lines with fixed values of the coupling of inflaton-\bsm~particle interaction (inclined lines) on $(\D\neff,\lh)$ plane are shown in Fig.~\ref{Fig:DNeff-vs-L12}. This figure shows that for a certain value of $\lpx$, $\D\neff$ grows as the value of $\lh$ decreases. A lower value of $\lh$ suggests less production of \sm~relativistic particle, and thus the ration in Eq.~\eqref{Eq:DNeff-for-our-case} increases.  
Fig.~\ref{Fig:DNeff-vs-L12} also gives a comparative view how the range of $\lpx$ varies for a given $\lh$ $(10^{-5}\geq \lh\geq 10^{-10} )$ and for the same range of $\D\neff$ $(0.29>\D\neff \geq 0.027)$ for different possible interactions of $X$ with $\phi$. 
For example, the required ranges for $\lpx$ for different interaction channels are - $10^{-11}\lsim y_\chi\lsim 10^{-6}$, $10^{-9}\lsim \lambda_{13,\varphi}\lsim 10^{-4}$.
 Additionally, this figure illustrates the allowed range of the value of the coupling $\lh$ regarding the \Planck~bound on $\D\neff$ for a given value of $\lpx$. 
 Moreover, for a given value of $\lpx$, this figure shows the value of $\lh$ that are in compliance with  \Planck~bound on $\D\neff$. For instance, the value $\lh$ should be $<3.26\times 10^{-9}$ for $y_\chi= 10^{-9}$. Furthermore, 
 the dashed-magenta colored line correspond to $\D\neff=0.254$, the best fit value predicted by Ref.~\cite{Guo:2017qjt}. The point where this line intersects with the inclined lines in that figure, gives the values of the coupling-pairs $(\lh,\lpx)$ correspond to the bound on $(n_s,r)$ plane when $\neff$ used as a free parameter in Fig.~\ref{Fig:Planck15+BK15-inflation-ns-r-bound}.

%
%

\subsection{Reheat temperature}

For the values of couplings shown in Fig.~\ref{Fig:DNeff-vs-L12}, the order of $\Trh/\sqrt{\mpl\,\mphi}\sim 7.60\times 10^{-10}$ (for $\pcc$  with $B_\chi\sim 0.02, \, y_\chi\sim 10^{-9},\, \lh\sim 6.97 \times 10^{-9}$), $\Trh/\sqrt{\mpl\,\mphi}\sim 10.64\times 10^{-10}$
(for $\phi\to \ug \ug$ with $B_\ug\sim 0.021, \, g_{\vp\ug}\sim 10^{-9},\, \lh\sim 9.76\times 10^{-9}$), $\Trh/\sqrt{\mpl\,\mphi}\sim 12.22\times 10^{-10}$
(for $\phi\to \vp\vp\vp$ with $B_\vp\sim 0.021, \, \lambda_{13,\vp}\sim 10^{-7},\, \lh\sim 1.12\times 10^{-7}$). All these values correspond to $\D\neff=\cmbs$ which can be measured by \cmbsfour. %
For the values of couplings shown in Fig.~\ref{Fig:DNeff-vs-L12}, the order of $\Trh/\sqrt{\mpl\,\mphi}$ and $\nrh$ are shown in Table.~\ref{Table:TRH} and it shows that $\Trh$ is well above Big Bang
Nucleosynthesis (\bbn) temperature ($\sim 1 \unit{MeV}$) for our chosen range of coupling values. 
\begin{table}[H]
\centering
\caption{\it Benchmark value for inflationary models (using $\ln(10^{10} A_s)=3.047\pm 0.014$ from $68\%$ TT,TE,EE+lowE+lensing+BAO~\cite{Planck:2018vyg} ).}
\label{Table:Benchmark-Inflation-Models}
\begin{tabular}{|c | c | c | c | c | c | c | c| c|} 
 \hline
 Inflation &  & $\frac{\phi^*}{\mpl}$ & $\frac{\phi_{\rm end}}{\mpl}$ & $\ncmb$ & $n_s$ & $r$ &   & $V(\phi_{\rm end})/\mpl^4$\\ 
Model &  &   &  &  &   &  &  & $ $ \\
 \hline
 S-I & - & $5.45$ & $0.94$ & $60$ & $0.9678$ & $0.00296$ & $\frac{\L}{\mpl}=3.1\times 10^{-3}$ & $2.7\times 10^{-11}$\\ \hline
N-I & $\frac{f_a}{\mpl}=4$ & $1.21$  & $11.17$ & $60$ & $0.9346$ & $0.012$ & $\frac{\L_N}{\mpl}=3.7\times 10^{-3}$ & $1.13\times 10^{-11}$\\
 \hline
 H-I & $\frac{v}{\mpl}=10$ & $3.88$ & $9.7$ & $60$ & $0.961$ & $0.0045$ & $\frac{\L_H}{\mpl}=3.5\times 10^{-3}$ & $1.64\times 10^{-11}$\\
 \hline
C-I & $\frac{f}{\mpl}=0.9$ & $0.009$ & $0.08$ & $60$ & $0.948$ & $5.7\times 10^{-8}$ & $A=1.1\times 10^{-14}$ & $1.79\times 10^{-15}$ \\ 
 \hline
\end{tabular}
\end{table}
\begin{table}[H]
\centering
\caption{\it Estimation of $\Trh$ and $\nrh$ for different interaction with $\phi$ and for four inflationary models with benchmark value shown in Table~\ref{Table:Benchmark-Inflation-Models}. For estimation of $\nrh$, we use $\mphi=10^{15}\GeV$.  ($\D\neff=0.06$ correspond to future reach from PICO/\cmbsfour).}
\label{Table:TRH}
\begin{tabular}{| c | c | c |c| c |c | c| c| c| c|} 
 \hline
  $\D\neff$ & $\phi\to X$ & $\lpx$ & $\lh$ & $\bx$ & $\lt(\Trh/\sqrt{\mpl}\rt)10^{10}$  & \multicolumn{4}{|c|}{$\nrh$ (Eq.~\eqref{Eq:efold-during-reheating})}\\
  \cline{7-10} 
  & & & &\eqref{Eq:BX-Decay-To-Fermion},\eqref{Eq:BX-Decay-To-Gauge-Boson}-\eqref{Eq:BX-Cubic-Boson} & (Eq.~\eqref{Eq:TRH}) & S-I & N-I & H-I  & C-I \\
 \hline
 \multirow{3}{4em}{$0.12$} & $\pcc$ & $y_\chi=10^{-9}$ & $6.97\times 10^{-9}$ & $0.02$ & $7.60\, \sqrt{\mphi}$ & $23.98 $ & $ 23.69$ & $23.82 $  & $20.78 $ \\
  \cline{2-10}
 %
  \cline{2-10}
 &$\phi\to \ug\ug$ & $g_{\phi\ug}=10^{-9}$ & $9.76\times10^{-9}$ & $0.021$ & $10.64 \,\sqrt{\mphi}$ & $23.54 $ & $ 23.25$ & $23.37 $  & $20.33 $\\
  \cline{2-10}
 &$\phi\to \vp\vp\vp$ & $\lambda_{13,\vp}=10^{-7}$ & $1.12\times10^{-8}$ & $0.021$ & $12.22 \,\sqrt{\mphi}$ & $23.35 $ & $ 23.06$ & $23.18 $  & $20.14 $ \\
 %
 \hline
\end{tabular}
\end{table}
%


\medskip

\section{Discussion and Conclusion}
\label{Sec:Discussion and Conclusion}
We assumed that the decay of inflaton during reheating epoch was the sole source of the production of an additional relativistic, free-streaming, non-self-interacting \bsm~particle contributing solely to $\neff$ in the form of \dr~and used the \cmb~data to constrain the branching fraction for its production. This gives us novel constraints (from \Planck) on the parameter space involving couplings and masses of the inflaton. Moreover we explore the detectable prospects that will be within the reach of next-generation \cmb~experiments. We summarize our main findings below:

\begin{itemize}

    \item 
   Fig.~\ref{Fig:Planck15+BK15-inflation-ns-r-bound} depicts that the best-fit values of $n_s$ and $r$, obtained from the numerical simulation of \Planck2015+\BICEP2 \cmb~data, alter depending on whether $\neff$ is fixed to its standard value or it is regarded as a variable, which thus also affects the selection of preferred inflationary models.
   While predictions regarding the values of $n_s$ and $r$ from three inflationary models - N-I, H-I, S-I are within $2\s$ contour from \Planck2015-\BICEP2 data with $\D\neff=0$, as depicted in Fig.~\ref{Fig:Planck15+BK15-inflation-ns-r-bound}, only predictions from H-I and S-I are within $2-\s$ bound when $\D\neff$ is allowed to vary to $0.254$.

    \item We studied the simplest scenario in which a relativistic non-thermal \bsm~particle, $X$, acts as the only source contributing to $\D\neff$. If $X$, together with \sm~relativistic particles, is produced from the decay of inflaton $\phi$, with branching fraction for the production of $X$ be given by $B_X$, as defined in Eq.~\eqref{Eq:definition-of-BX},
    then the contribution of $X$ to $\D\Neff$, as defined in Eq.~\eqref{Eq:DNeff-for-our-case}, is solely determined by $B_X$ and independent of $\G_\phi$. 
    Additionally, as shown in  Fig.~\ref{Fig:Branching-fraction-and-bound-on-DNeff}, $\D\neff$ is a monotonically increasing linear function of $\bx$. This is because a greater value of $\bx$ implies a larger generation of $X$ particles. 
    Furthermore, \Planck+BAO bound on $\neff$ has already eliminated the possibility of $B_X > 0.09$, and $\bx$ within the range $0.09\gsim B_X \gsim 0.009$ can be testified by future \cmb~observations (e.g. CMB-HD) with better sensitivity for $\Neff$.

   \item 
   Using the combination of $(r, \bx)$ best-fit contour (transformed from $(r, \D\neff)$ best-fit contour from \cmb~data), the predicted value of $r$ from inflationary models, and \cmb~bounds on $\neff$, we showed in Fig.~\ref{Fig:r-Br-Bound-region-plot} that we can conclude whether $X$ can be completely produced from the decay of inflaton or whether that inflationary model remains a preferred one if $X$ is entirely produced from inflaton decay. When inflaton acts the source of $X$, then Fig.~\ref{Fig:r-Br-Bound-region-plot} illustrates that four inflationary models will be within $2\s$ contour of `TTTEEE+tau' data if $\bx \gsim 0.033$ (for N-I), $\bx\gsim 0.061$ (for H-I), $\bx\gsim 0.097$ (for C-I), $ 0.11$ (for S-I). Similarly, to be within $2\s$ contour of `TTTEEE+lowTEB+BKP' it is required that $\bx\gsim 0.046$ (for N-I), $\bx\gsim0.07$  (for H-I), $\bx\gsim 0.13$  (for C-I), $\bx\gsim 0.15$ (for S-I). This conclusion, along with the Fig.~\ref{Fig:Branching-fraction-and-bound-on-DNeff}, indicates that the assumption that inflaton as the sole source of $X$ which is the only particle that is contributing completely to $\D\neff$, is not compatible for S-I and C-I (regarding `TTTEEE+tau' data). However, for large field C-I, it is possible for the inflaton to be the sole source of $X$ which is the only particle that is contributing completely to $\D\neff$, is compatible. For the regularized form of H-I, the conclusion remains the same as for H-I, albeit the lower limit of $\bx$ is $\gsim 0.02$.

\item 
We identified the permissible region on two-dimensional $(\lpx,\lh)$ plane in Figs.~\ref{Fig:Allowed-region-Plane-of-TwoCouplings} for all the interaction channels mentioned in Eqs.~\eqref{Eq:DecayToFermion}-\eqref{Eq:DecayToGaugeBoson}, surviving bounds on $\neff$ from the current \cmb~observations and also the parameter-space accessible to the future \cmb~observations listed in Table~\ref{Table:bound-on-Delta-Neff}.
The permissible region on $(y_\chi,\lh)$ plane for $\phi\to \tilde{\chi}\chi$ is identical to that on $(\lambda_{12,\varphi},\lh)$ plane for $\phi\to \vp\vp$ when $\s^\prime_m=\mphi$, although a lower range for $(\lambda_{12,\varphi},\lh)$ is required for non-thermal \dr~scenario. 
To remain within \Planck~bound, $y_\chi/\lh\lsim 0.32$, $\lambda_{12,\vp}/\lh \lsim 3.18 \, (\text{when }\s^\prime_m=0.1\mphi)$, $\lambda_{12,\vp}/\lh \lsim 0.032 \, (\text{when }\s^\prime_m=10\mphi)$, $\lambda_{13,\varphi}/\lh \lsim 19.37$,
$g_{\phi\ug}/\lh \lsim 0.022\, (\text{when }\L_m=0.1\mphi),\, \lsim  0.23 \,  (\text{when }\L_m=\mphi), \lsim 2.28\,  (\text{when }\L_m=10\mphi)$. 
When $\s^\prime_m$ is raised, the allowable area on the $(\lambda_{12,\vp}, \lh)$ plane shrinks. In contrast, as $\L_m$ grows, so does the permissible area on the $(g_{\phi\ug}, \lh)$ plane (see Fig.~\ref{Fig:mass-scale-varied}).

\item Fig.~\ref{Fig:DNeff-vs-L12} depicts the variation of the range of $\lpx$ for the same range of $\lh$ ($10^{-10}\lsim \lh \lsim 10^{-5}$) and same range of $\D\neff$ ($0.29\gsim \D\neff \gsim0.027$): $ 10^{-6}\gsim y_\chi \gsim 10^{-11}$ (the range of $g_{\phi\ug}$ is nearly identical to the range of $y_\chi$ ), $10^{-4}\gsim \lambda_{13, \vp} \gsim 10^{-9})$. 
Forbye, Fig.~\ref{Fig:DNeff-vs-L12} shows for a given value of $\lpx$, the lower limit of $\lh$ that is already eliminated by \Planck+BAO and the order of the value of $\lh$ that can be tested by CMB-HD or other \cmb~observations. For instance, when $\chi$ is the \bsm, for $y_\chi\sim 10^{-8}$, $\lh> 3\times 10^{-8}$ is already excluded by \Planck, and $\lh\gsim  10^{-7}$ can be examined by future \cmb~observations.

\item  Fig.~\ref{Fig:DNeff-vs-L12} gives the values of the  pairings $(\lpx, \lh)$ for which $\D\neff=0.254$. 
For these values of the pairings $(\lpx, \lh)$, S-I model will be within $2\s$ contour and H-I will be within $1\s$ contour-bound on the $(n_s, r)$ plane  when $\D\neff$ is treated as a variable rather than being forced to be $0$, as shown in Fig.~\ref{Fig:Planck15+BK15-inflation-ns-r-bound}.

\item For the values of couplings shown in Fig.~\ref{Fig:DNeff-vs-L12}, the order of $\Trh/\sqrt{\mphi}\sim 1.18 {\sqrt{\GeV}}$ (for $\pcc$ with $ y_\chi\sim 10^{-9},\, \lh\sim 6.97 \times 10^{-9}$), $\Trh/\sqrt{\mphi}\sim 1.65{\sqrt{GeV}}$
(for $\phi\to \ug \ug$ with $ g_{\vp\ug}\sim 10^{-9},\, \lh\sim 9.76\times 10^{-9}$), and $\Trh/\sqrt{\mphi}\sim 1.89{\sqrt{GeV}}$
(for $\phi\to \vp\vp\vp$ with $\lambda_{13,\vp}\sim 10^{-7},\, \lh\sim 1.12\times 10^{-8}$). All these values correspond to $B_X\sim 0.02$ or $\D\neff=\cmbs$ which can be measured by the upcoming \cmbsfour/PICO~experiment.
\end{itemize}

In future, we may be able to extend our analysis to scenarios in which the \dr~from inflaton decay may alleviate the $H_0$ tension caused by the presence of extra \dr, and how this may have profound implications for inflation model selection, but this is beyond the scope of the current manuscript and will be pursued in a subsequent publication.

\medskip

\section*{Acknowledgement}
Work of Shiladitya Porey is funded by RSF Grant 19-42-02004. 

\appendix

\section{}
\begin{figure}[H]
    \centering
    \includegraphics[width=0.75\linewidth]{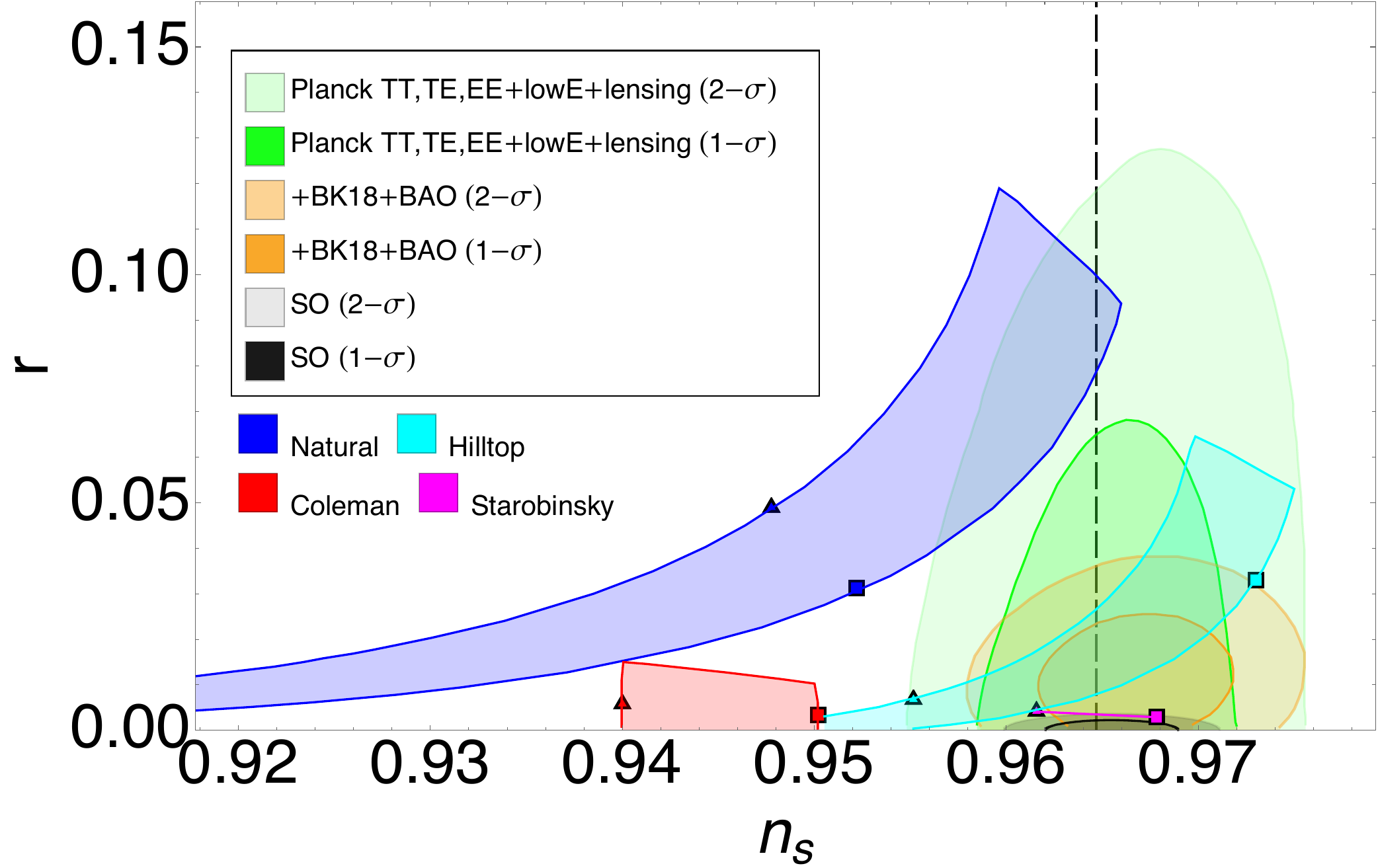}
    \caption{\it \raggedright \label{Fig:[Newest]inflation-ns-r-bound} 
    Illustration of $n_s-r$ predictions for four inflationary models together with current $1-\sigma$ and $2-\sigma$ best-fit contour from \Planck2018 (green shaded region), \Planck2018+\BICEP3+\KeckArray2018 (yellow shaded region) and future \cmb~observation (SO) with black shaded region.
}  
\end{figure}

\label[appendix]{NewBound}
Older version of \Planck, \BICEP~data (\Planck2015, \BICEP2) have been used in Fig.~\ref{Fig:Planck15+BK15-inflation-ns-r-bound} to draw 2D posterior constraints with standard value of $\Neff$, in the $(n_s,r)$ plane with the $95\%$ and $68\%$ CL contour bounds. In Fig.~\ref{Fig:[Newest]inflation-ns-r-bound}, the most recent 
bounds from different \cmb~observations (\Planck2018+\BICEP3+\KeckArray2018) along with bounds from future \cmb~observation The Simons Observatory (SO) 
are put on view. The black dashed vertical line corresponds to $n_s=0.9647$~\cite{Aghanim:2018eyx}. Along with that, $n_s-r$ prediction from four single-field slow roll inflationary models (which have been already presented in Fig.~\ref{Fig:Planck15+BK15-inflation-ns-r-bound}) are also shown in Fig.~\ref{Fig:[Newest]inflation-ns-r-bound}. %
Among the four considered models,
two of them have been disfavored even by $2-\sigma$ contour, while both S-I and H-I predict values of $n_s,r$ that still fall within $1-\sigma$ best-fit contour of \Planck2018+\BICEP3+\KeckArray2018. However, only S-I is favored by the analysis of the SO observation, but limited to $\ncmb\sim 60$.

\section{}
\label[appendix]{Linde's section}
As mentioned previously, the potential for H-I (Eq.~\eqref{eq:Hilltop-potential}) is not bounded from below. Introducing other term to the potential to stabilize the potential of Eq.~\eqref{eq:Hilltop-potential} affects the $n_s,r$ predictions of H-I
except for the scenario where 
$v<\mpl$, where additional term $\propto \phi/v$ can be introduced in the potential . Nevertheless, for such scenario, predicted value of $n_s$ even for $\ncmb=60$ is less ($n_s=0.95$) compared to best-fit $1-\s$ contour from \cmb~data~\cite{Kallosh:2019jnl,Kallosh:2019hzo,Kallosh:2021mnu}.  To resolve these issues, the regularized form of the H-I (although this form of hilltop potential lacks significant theoretical motivation in contrast to the form of Eq.~\eqref{eq:Hilltop-potential}) has been suggested for $v\gsim\mpl$ as~\cite{Kallosh:2019jnl}
\eq{\label{Eq:regularizedHilltop}
V(\phi)=\L_H^4 \qty[1- \qty(\frac{\phi}{v})^4]^2
\,.
}

\begin{figure}[H]
    \centering
    \includegraphics[width=0.48\linewidth]{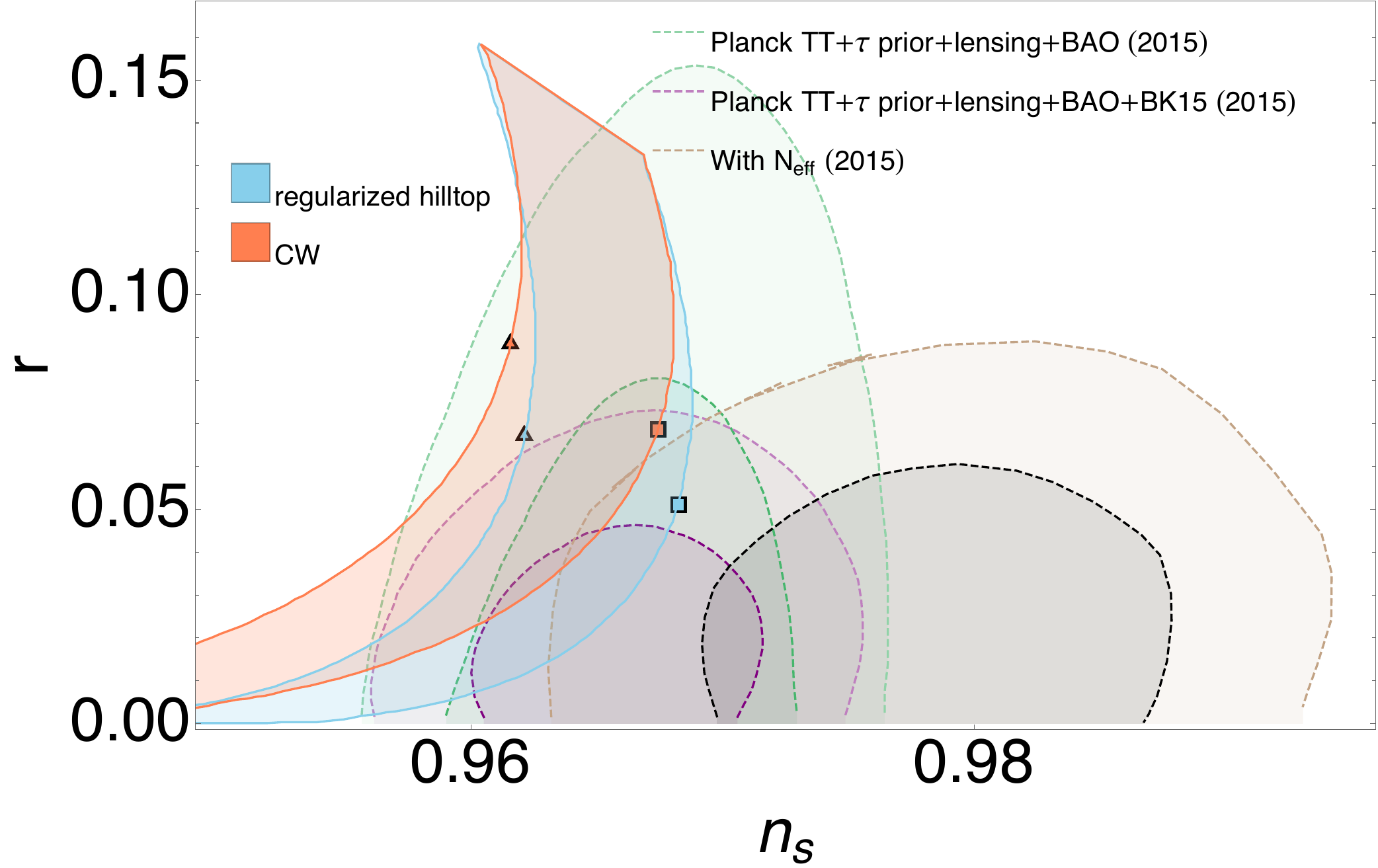}
    \includegraphics[width=0.48\linewidth]{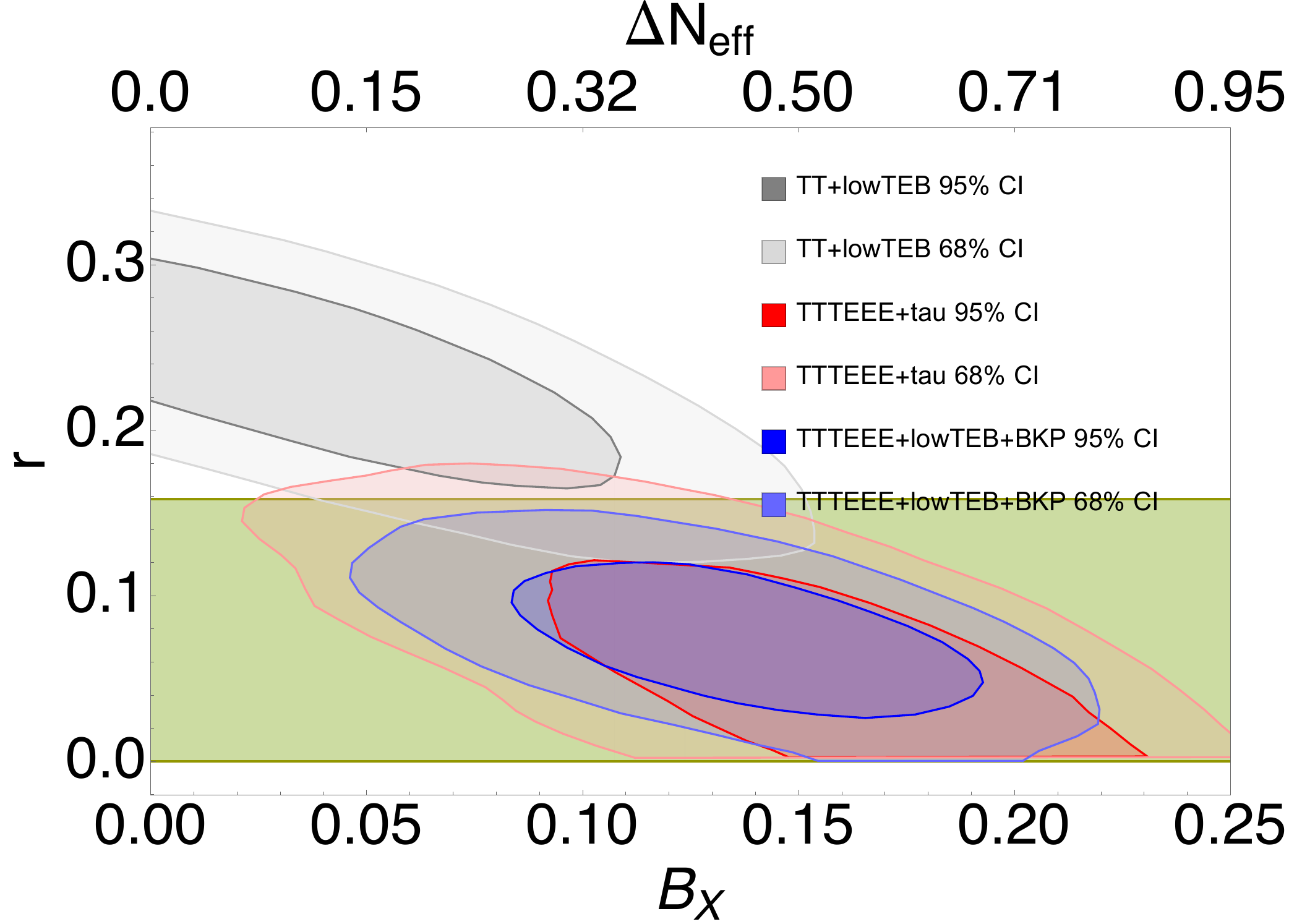}
    \caption{\it \raggedright \label{Fig:RegularizedHilltop-CW-Linde} 
    {\bf Left panel: }illustration of $n_s-r$ predictions for regularized Hilltop inflation (Eq.~\eqref{Eq:regularizedHilltop}) and C-I inflationary model (Eq.~\eqref{Eq:CW-I-model}) but for $f\gsim \mpl$ together with $1-\sigma$ and $2-\sigma$ contour from \Planck2015, and \Planck2015+\BICEP2-\KeckArray2015 and bound from analysis where $\D\neff$ is treated as a variable from~\cite{Guo:2017qjt,Kannike:2018zwn}, as described in Fig.~\ref{Fig:Planck15+BK15-inflation-ns-r-bound}. {\bf Right panel:} olive-green shaded region shows the allowed region of $r$ for regularized Hilltop/ C-I (with $f\gsim \mpl$) inflationary models. Additionally, the best fit contour on $(r, \bx)$ plane in the background is also depicted from Ref.~\cite{DiValentino:2016ucb}, as mentioned in Fig.~\ref{Fig:r-Br-Bound-region-plot}.
}  
\end{figure}

We have considered $n_s,r$ predictions of Regularized Hilltop inflation and also C-I for $m\gsim\mpl$ for the completeness of our discussion, and displayed it in the left panel of Fig.~\ref{Fig:RegularizedHilltop-CW-Linde} as cyan-colored and red-colored regions for $\ncmb=50-60$ (line with a triangle and line with a square correspond to $\ncmb=50$ and $\ncmb=60$). Both of these inflationary models (regularized Hilltop and CW for $f>\mpl$) predicts $n_s,r$ values even within $1-\s$ best-fit contour from the analysis of \Planck2015+\BICEP2-\KeckArray2015. Interestingly, these two inflationary models also predict $n_s, r$ values that fall inside $2-\s$ CL of the best-fit contour from~\cite{Guo:2017qjt} with best-fit value of $\D\neff=0.254$. The maximum value of $r$ predicted by regularized Hilltop inflationary model used on the left panel is $r=0.158$ (for $\ncmb=50$) which is also the same as the maximum value of $r$ considered for CW inflation. The horizontal olive-colored stripe on the right panel of~\cref{Fig:RegularizedHilltop-CW-Linde} indicates the range of the value of $r\lsim 0.158$. The $r-B_X$ best-fit contours on the background of the right panel of~\cref{Fig:RegularizedHilltop-CW-Linde} are from Ref.~\cite{DiValentino:2016ucb}, as depicted already in~\cref{Fig:r-Br-Bound-region-plot}. Since $B_X\gsim 0.02$ and $B_X\gsim 0.05$ is needed for both regularized Hilltop and C-I inflation to be within at least $68\%$ of CL`TTTEEE+tau' and `TTTEEE+lowTEB+BKP' datasets, we conclude that $X$ produced from the decay of inflaton during reheating and contributing solely and completely to $\D\neff$ are simultaneously compatible for these two inflationary scenarios.  

\section{}
\begin{figure}[H]
\centering
\includegraphics[height=6cm,width=8cm]{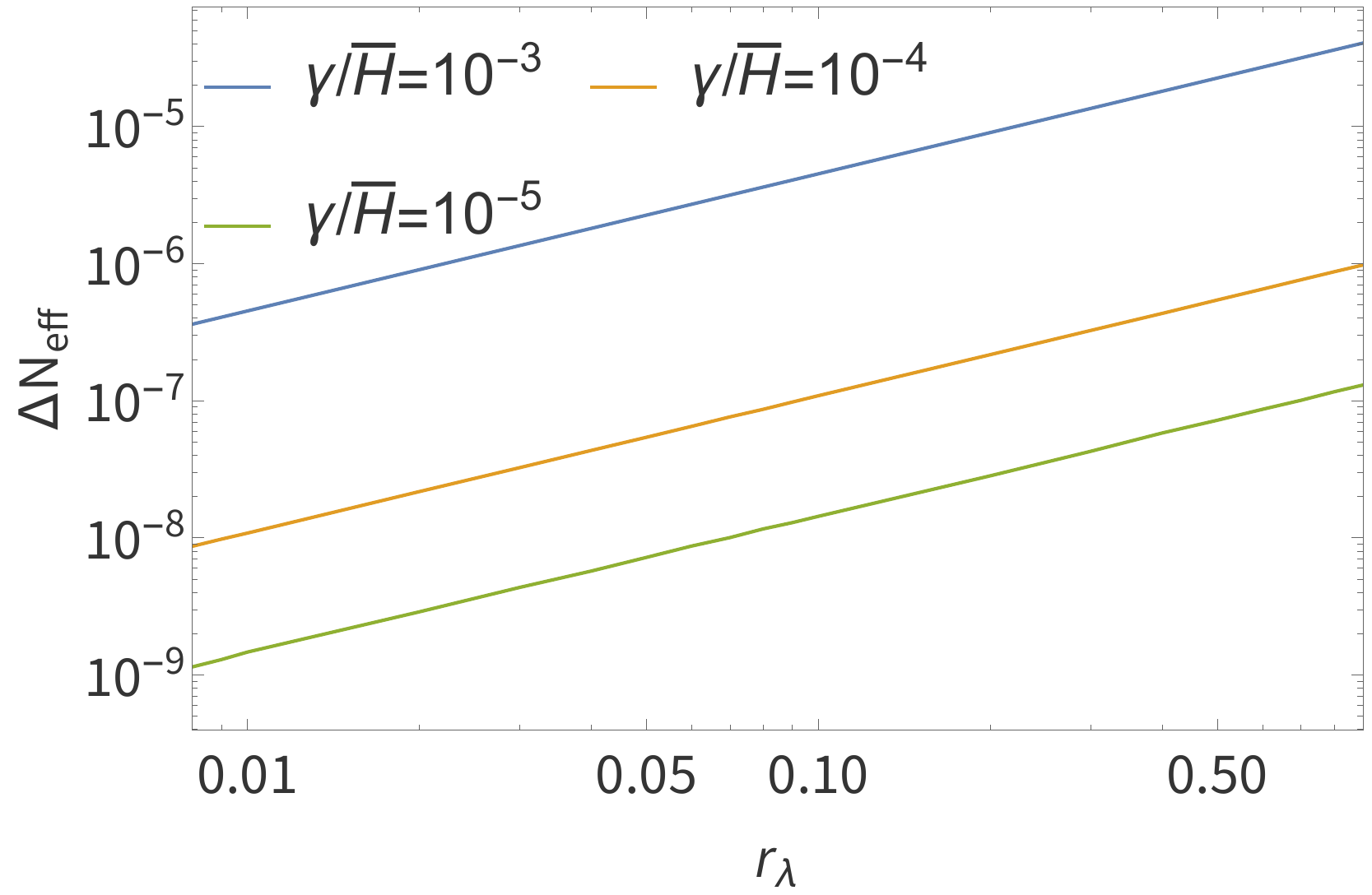}
\caption{\it \raggedright 
{
Variation of $\D\Neff$ versus $r_\lambda$ for three different values of $\gamma/\hs$ for a scalar \bsm~particle ($\varphi$) contributing to $\D\neff$ and produced via~\cref{Eq:ScatteringToBoson}. Contribution to $\D\neff$ decreases with smaller values of both $\gamma/\hs$ and $r_\lambda$. }
}
\label{Fig:ScalarDRScatteringNeff}
\end{figure}

\label[appendix]{Sec:Boltzman solve for scalar DR scattering}
Here, we consider that $X$ is a scalar particle $\varphi$ which is produced through $\phi \phi \to \varphi\varphi$ interaction channel.
Accordingly, based on~\cref{Eq:ScatteringToBoson}, the reaction rate between $\phi$ and $X$-particles can be written as~\cite{Drewes:2019rxn,Garcia:2020wiy}
\begin{align}
    \G_{\phi\phi\to \varphi\varphi}=\frac{\lambda_{22,\varphi}^2 \, \lt<\phi\rt>^2}{256 \pi\, \mphi} 
    =\frac{\lambda_{22,\varphi}^2 \, \rho_\phi}{128 \pi\, \mphi^3} \,, \qquad
&\lt(\text{If}\,X\equiv\varphi, \,\text{and}\,   \phi\phi \to \varphi\varphi\rt)\,.    \label{Eq:BX-Scattering-To-Boson}
\end{align}
Then %
    $\G_\phi=\G_{\phi \to hh}+\G_{\phi\phi\to \varphi\varphi}$.
Since we consider the scenario where the potential of inflaton about the minimum is quadratic during reheating, $m_\phi$ is independent of time. However, $\rho_\phi$ and, consequently, $\G_{\phi\phi\to \varphi\varphi}$, both evolve with time~\cite{Garcia:2020wiy}. The Boltzmann equations (\cref{Eq:BG1,Eq:BG2,Eq:BG3}) and Friedman equation (\cref{eq_H}) can be expressed using rescaled dimensionless variables as follows 
\begin{align}
  & \frac{\dd P}{\dd \tau} + 3 \,\hs \,P = -\gamma \left( 1+ r_\lambda \, R\right) \,P\,, 
  \label{Eq:BG1-1}\\  
  & \frac{\dd R}{\dd \tau} + 4 \, \hs\,R = \gamma\,P\,,  \label{Eq:BG2-2}\\
  & \frac{\dd {\cal X}}{\dd \tau} + 4 \,\hs\, {\cal X} = \gamma\, r_\lambda\, R \,P\,,
  \label{Eq:BG3-3}\\
  &\hs^2= P+R+{\cal X}\,,\label{eq_H-H}
\end{align}
where $\tau = t \, {\cal H}_I\,, \hs= {\cal H}/{{\cal H}_I}\,, {\cal H}_I^2 =\qty( {1}/{3 M_P^2}) \Phi_I\,, P={\rho_\phi}/{\Phi_I}\,, R ={\rho_{\rm rad}}/{\Phi_I}\,, {\cal X}={\rho_X}/{\Phi_I}\,, \gamma= {\Gamma_{\phi\to {h h}}}/{{\cal H}_I}$. Here, we assume that ${\rho_{\rm rad}}=\r_X=0$ at the onset of the reheating era, with $\Phi_I$ denoting the energy density of inflation and ${{\cal H}_I}$  representing the value of Hubble parameter at the beginning of reheating. The dimensionless parameter $r_\lambda$ is defined as
\begin{align}
   r_\lambda= \frac{1}{16 \, m_\phi^4}\frac{\lambda_{22,\varphi}^2 }{\lambda_{12,H}^2} \, \Phi_I\,.
\end{align}
In this work, we assume $r_\lambda<1$ to avoid the universe being dominated by dark sector particles. By solving~\cref{Eq:BG1-1,Eq:BG2-2,Eq:BG3-3,eq_H-H} and substituting the results into~\cref{Eq:DNeff-for-our-case} we obtain the variation of $\D\Neff$ with respect to $r_\lambda$, considering three different values of $\gamma/\hs$, as depicted in Fig.~\cref{Fig:ScalarDRScatteringNeff}. From~\cref{Fig:ScalarDRScatteringNeff}, we conclude that the contribution of scalar \bsm~particles in $\D\neff$ is very small when $\varphi$ is produced via~\cref{Eq:ScatteringToBoson} rather than~\cref{Eq:DecayToBoson}. This is because, as the reheating phase progresses, the reaction rate, and therefore $\G_{\phi\phi\to \varphi\varphi}/\G_{\phi \to hh}$, decreases with $\rho_\phi$. Due to the same reason, contribution to $\D\neff$ in this scenario ($\phi\phi \to \varphi\varphi$) depends on $\gamma/\hs$ in contrary to $\phi \to \varphi\varphi$ or other interaction channels mentioned earlier (\cref{{Eq:DecayToFermion},Eq:QuibicInteractionToBoson,Eq:DecayToGaugeBoson}). The decrease of $\D\neff$ for smaller values of $r_\lambda$ is attributed to the same reason observed in the decrease of $\D\neff$ with $\bx$ in~\cref{Fig:Branching-fraction-and-bound-on-DNeff},  which is  smaller fraction of $\rho_\phi$ being converted to $\rho_X$ with smaller values of $r_\lambda$.


\section{}
\begin{figure}[H]
    \centering
    \includegraphics[width=0.5\linewidth]{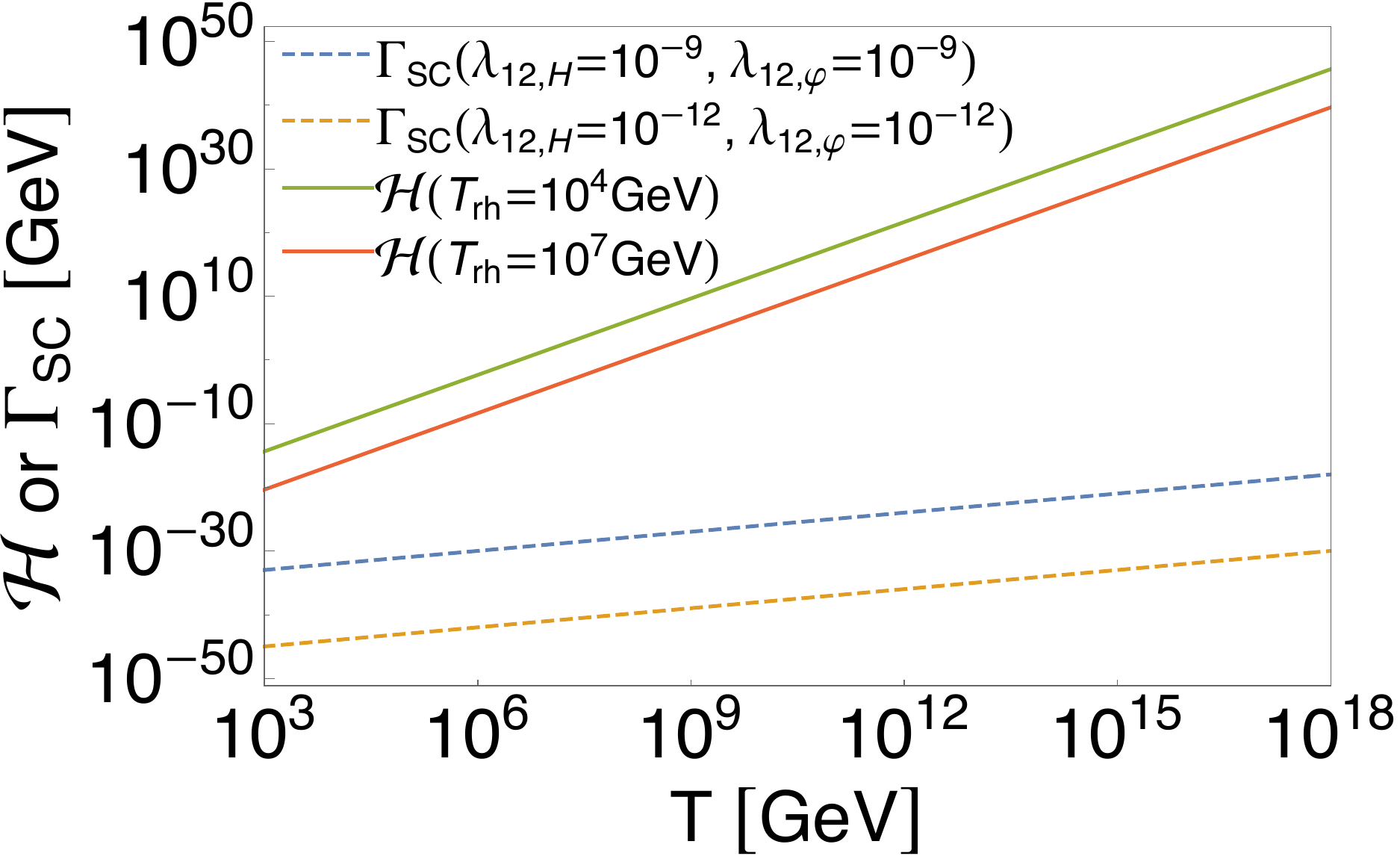}
    \caption{\it \raggedright
    Presentation of representative sets of values for the couplings $(\lambda_{12,H},\lambda_{12,\varphi})$ mentioned in~\cref{Eq:Inflaton decaying to two higgs,Eq:DecayToBoson}, such that scalar \dr~particles cannot achieve thermal equilibrium with the \sm~Higgs via inflaton exchange diagram at the tree level. The solid lines correspond to ${\cal H}$ from~\cref{Eq:Hubble during reheating} for two different values of $\Trh$, while the dashed lines correspond to $\Gamma_{SC}$ from~\cref{eq:scalar-dr--scat-Gamma},  for different values of $(\lambda_{12,H},\lambda_{12,\varphi})$.
    %
    %
    }
    \label{fig:scalarDR-vs-Hubble}
\end{figure}

\label[appendix]{Sec:thermal equilibrium fish}
Here, we discuss the conditions, e.g. ranges of the couplings mentioned in~\cref{Eq:BSM+Higgs-production-lagrangian}, for which our assumption that \dr~particles are not thermalized with the radiation bath, remains valid. More precisely, we explore the conditions wherein these particles cannot attain kinetic equilibrium%
\footnote{When both \sm~Higgs and \dr~particles are relativistic, we can assume that their chemical potential is negligible. Hence, kinetic equilibrium implies thermal equilibrium during that epoch.}
with \sm~Higgs via inflaton exchange diagrams. Let us begin by considering 2-to-2 scattering of \dr~particles with inflaton as the mediator. 
The reaction rate (per Higgs particle) is given by
\begin{align}
    \Gamma_{SC}= \expval{\sigma \, \abs{v_{\rm rel}}} \, n_{\dr}=\sigma \, \abs{v_{\rm rel}}\, n_{\dr}\,,
\end{align}
where $n_{\dr}$ is the number density of \dr~particles, $\abs{v_{\rm rel}}$ represents the absolute value of the relative velocity between projectile and target particles, and the brackets denote averaging over the phase space. The rightmost equality holds based on our assumption that cross-section $\sigma$ is independent of $v_{\rm rel}$. During the reheating era, we can assume $T\gg M_H$, where $M_H$ is the mass ($\approx 125 \GeV$) of \sm~Higgs particles. 
Therefore, thermal Higgs particles are relativistic at that time  with energy $\approx T$. Furthermore, \dr~particles are expected to possess considerably low masses 
to contribute to $\D\Neff$. We can also assume that \dr~particles produced from the decay of inflaton, are also relativistic during the epoch we are interested.
Since, \dr~particles are non-thermal, their temperature need not necessarily 
align with 
that of \sm~Higgs i.e. $T$. Nevertheless, we can approximate that energy of \dr~particles $\sim {\cal O}\qty(T)$. Therefore, 
$\abs{v_{\rm rel}}\sim 1$ and $n_{\dr}\sim T^3$~\cite{Dolgov:2009zj}. 
For 2-to-2 scattering process, the total cross-section in center-of-mass frame
\begin{align}\label{eq:formula}
    \sigma \sim  \int \frac{\overline{\abs{{\cal M}}^2}}{(E_1+E_2)^2}  \frac{\abs{{\bf p}_f} }{\abs{{\bf p}_i}} \, \dd \Omega\,,
\end{align}
where $\qty(E_1+E_2)$ is the sum of the energies of the incoming particles, $\abs{{\bf p}_f}$ is the magnitude of the 3-momentum for either of the outgoing particles, and $\abs{{\bf p}_i}$ is the
magnitude of 3-momentum for either of the incoming particles, ${\cal M}$ stands for the Feynman amplitude with the over-line indicating an average taken over unmeasured 
spins or polarization states of the involved particles, and the integration is performed over solid angle $\Omega$. Following the aforementioned discussion, we assume $E_1,E_2\sim T$ and $\abs{{\bf p}_f}\sim \abs{{\bf p}_i}$. For the scenario, when \dr~particles are scalar, and interact with \sm~Higgs particles with inflaton as the mediator, we get at tree-level
\begin{align}
\overline{\abs{{\cal M}}^2} \sim \frac{1}{m_\phi^4}    \lt( \lambda_{12,H} \, \s_m\, \lambda_{12,\varphi}\, \s^\prime_m\rt)^2\,.
\end{align}
For $\s_m\,,\s^\prime_m\sim m_\phi$, and utilizing~\cref{eq:formula}, we get reaction rate
\begin{align}\label{eq:scalar-dr--scat-Gamma}
   \Gamma_{SC}\sim  \lt( \lambda_{12,H} \,  \lambda_{12,\varphi}\rt)^2 \, T\,.
\end{align}
During reheating era, Hubble parameter  can be defined as (such that at $T=\Trh$, $\rho_\phi=\rho_{\rm rad}$)~\cite{Garcia:2020eof,Mambrini:2021zpp}
\begin{align}
    {\cal H}\sim \frac{T^4}{M_P \Trh^2}\,.\label{Eq:Hubble during reheating}
\end{align}
Therefore,  if $\Gamma_{SC}< {\cal H}$ during reheating, \dr~cannot achieve thermal equilibrium with \sm~Higgs via inflaton exchange process. The values of $(\lambda_{12, H},\lambda_{12,\vp})$ for which $\Gamma_{SC}< {\cal H}$ is maintained, is shown in~\cref{fig:scalarDR-vs-Hubble}. The solid lines in~\cref{fig:scalarDR-vs-Hubble} represent ${\cal H}$ calculated from~\cref{Eq:Hubble during reheating} for two different values of $\Trh$: $\Trh\sim 10^{4}\GeV$ and $\Trh\sim 10^{7}\GeV$, and the dashed lines  correspond to $\Gamma_{SC}$ for  different values of $(\lambda_{12, H},\lambda_{12,\vp})$. 
From~\cref{fig:scalarDR-vs-Hubble} we conclude that even for $\Trh\sim 10^7\GeV$, along with $(\lambda_{12, H}\lsim 10^{-9},\lambda_{12,,\vp}\lsim 10^{-9})$ thermal equilibrium between scalar \dr~and \sm~Higgs particles (via inflaton exchange process) is not feasible. This justifies the range of $(\lambda_{12, H},\lambda_{12,\vp})$ we have considered (for instance, see~\cref{Fig:mass-scale-varied}). 
Reheating scenarios where $\Trh \gsim {\cal O}(10^7)\GeV$ may lead to an excessive production of gravitino~\cite{Kolb:2003ke}. Additionally, several proposed scenarios suggest the plausibility of $\Trh$ being $\sim 10^4 \GeV$ or even lower (for example, see~\Ccite{Ellis:2021kad,Hannestad:2004px}). 
For higher values of $\Trh$, we need to consider  smaller values of $(\lambda_{12, H},\lambda_{12,\vp})$ to ensure that scalar \dr~particles cannot reach thermal equilibrium with \sm~Higgs.  
For fermionic \dr~particles (see~\cref{Eq:DecayToFermion}) and $T\gg M_H$, the square of Feynman amplitude of the 2-to-2 scattering process between \dr~and \sm~Higgs particles via inflaton exchange diagram is 
\begin{align}
\overline{\abs{{\cal M}}^2} \sim \frac{1}{m_\phi^4}    \lt( \lambda_{12,H} \, \s_m\, y_\chi\rt)^2 \, \mathbb{f}_\chi\,,
\end{align}
where $\mathbb{f}_\chi\sim \sum_{\text{spins}}\lt[ \bar{u}_{\chi} v_{\bar{\chi}} \rt]\lt[\bar{u}_{\chi} v_{\bar{\chi}}\rt]^* \sim 4(p_\chi \cdot p_{\bar{\chi}}- m_\chi^2)\sim T^2$. Here, $u_{\chi}$, and $v_{\bar{\chi}}$ are Dirac spinors, and $p_\chi,  p_{\bar{\chi}}$ are 4-momenta of fermionic particles. This along with $\s_m\sim m_\phi$, leads to
\begin{align}\label{eq:fermion-dr--scat-Gamma}
   \Gamma_{SC} \sim  \frac{1}{m_\phi^2}    \lt( \lambda_{12,H} \,  y_\chi \rt)^2 \, T^3\,.
\end{align}
\begin{figure}[H]
    \centering
    \includegraphics[width=0.5\linewidth]{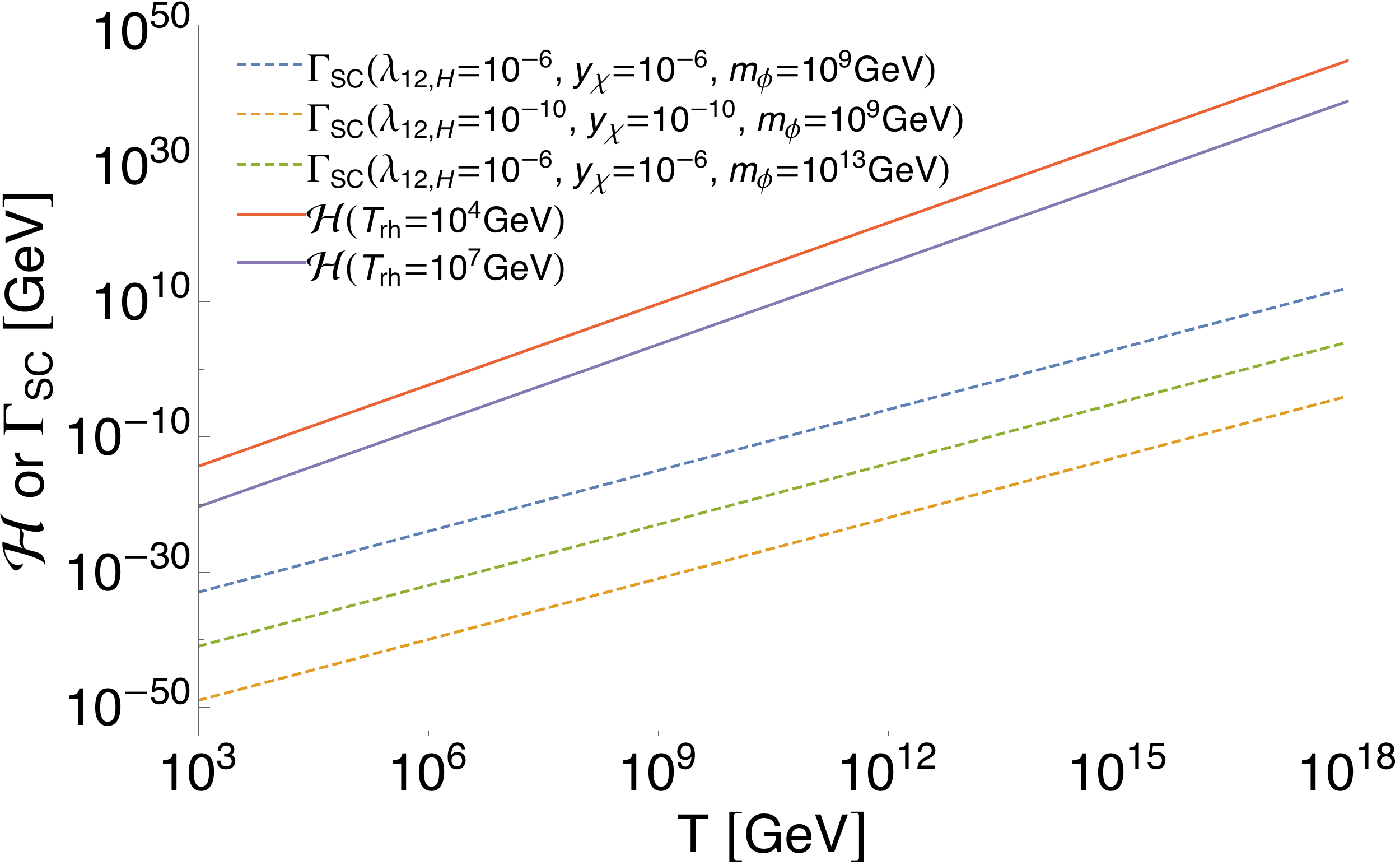}
    \caption{\it \raggedright 
    Presentation of representative sets of values for the couplings $(\lambda_{12,H},y_\chi)$ mentioned in~\cref{Eq:Inflaton decaying to two higgs,Eq:DecayToFermion}, such that fermionic \dr~particles cannot achieve thermal equilibrium with the \sm~Higgs via inflaton exchange diagram at the tree level. The solid lines correspond to ${\cal H}$ from~\cref{Eq:Hubble during reheating} for two different values of $\Trh$, while the dashed lines correspond to $\Gamma_{SC}$ from~\cref{eq:fermion-dr--scat-Gamma},  for different values of $(\lambda_{12,H},y_\chi)$ and $m_\phi$.
    %
    }
    \label{fig:fermionDR-vs-Hubble}
\end{figure}
\cref{fig:fermionDR-vs-Hubble} illustrates sample values of $(\lambda_{12, H},y_\chi)$ for which the condition $\Gamma_{SC}< {\cal H}$ holds.
From~\cref{fig:fermionDR-vs-Hubble}, we find that achieving thermal equilibrium between \dr~and \sm~Higgs particles is not possible for $\Trh\sim 10^7\GeV$, considering values of couplings $(\lambda_{12, H}\lsim 10^{-6},y_\chi\lsim 10^{-6})$, and $m_\phi\sim 10^9\GeV$. However, higher values of $\Trh$ necessitate either smaller values of $(\lambda_{12, H},y_\chi)$, or a larger value $m_\phi$, or both, to ensure that fermionic \dr~particles cannot reach in thermal equilibrium with \sm~Higgs via inflaton exchange process.

When \dr~particles are $U(1)$ gauge boson, and produced for the interaction in~\cref{Eq:DecayToGaugeBoson}, the square of Feynman amplitude of 2-to-2 scattering process between \dr~and \sm~Higgs particles via inflaton mediation, is given by
\begin{align}
\overline{\abs{{\cal M}}^2} \sim \frac{1}{T^2}    \lt( \frac{\lambda_{12,H} \, \s_m}{m_\phi^2}\, \frac{g_{\phi \ug}}{\L_m}\rt)^2 \, \mathbb{f}_{\ug}\,,
\end{align}
where $\mathbb{f}_{\ug}$ is $\sim(\text{momentum of DR particle})^4$. This can be explained as follows: 
If $A^\mu$ is the vector potential of $\ug$, and we write it in terms of annihilation ($a_{\lambda_p}(\mathbf{k})$) and creation operator $(a_{\lambda_p}^\dagger(\mathbf{k}))$, then $A^\mu$ contains a term $a_{\lambda_p}^\dagger(\mathbf{k})\, \epsilon_{\lambda_p}^{\mu *} \, e^{ikx}$, where the bold symbol denotes 3-vectors, $\epsilon_{\lambda_p}^{\mu}$ is the polarization vector, and $\lambda_p$ in the subscript is polarization index. Hence, $\partial_\mu A^\nu$ contains a term $k_\mu \, a_{\lambda_p}^\dagger(\mathbf{k})\, \epsilon_{\lambda_p}^{\nu *} \, e^{ikx}$. When we define $F_{\m\n}F^{\m \n}$ and then, calculate $\abs{\cal M}^2$ we get for our scenario $\mathbb{f}_{\ug}\sim T^4$.
%
%
Considering $\L_m, \s_m\sim m_\phi$, we get
\begin{align}\label{eq:gauge-dr--scat-Gamma}
   \Gamma_{SC} \sim \frac{1}{m_\phi^4} \lt(\lambda_{12,H} \,  g_{\phi \ug} \rt)\, T^5\,.
\end{align}
\begin{figure}
    \centering
    \includegraphics[width=0.5\linewidth]{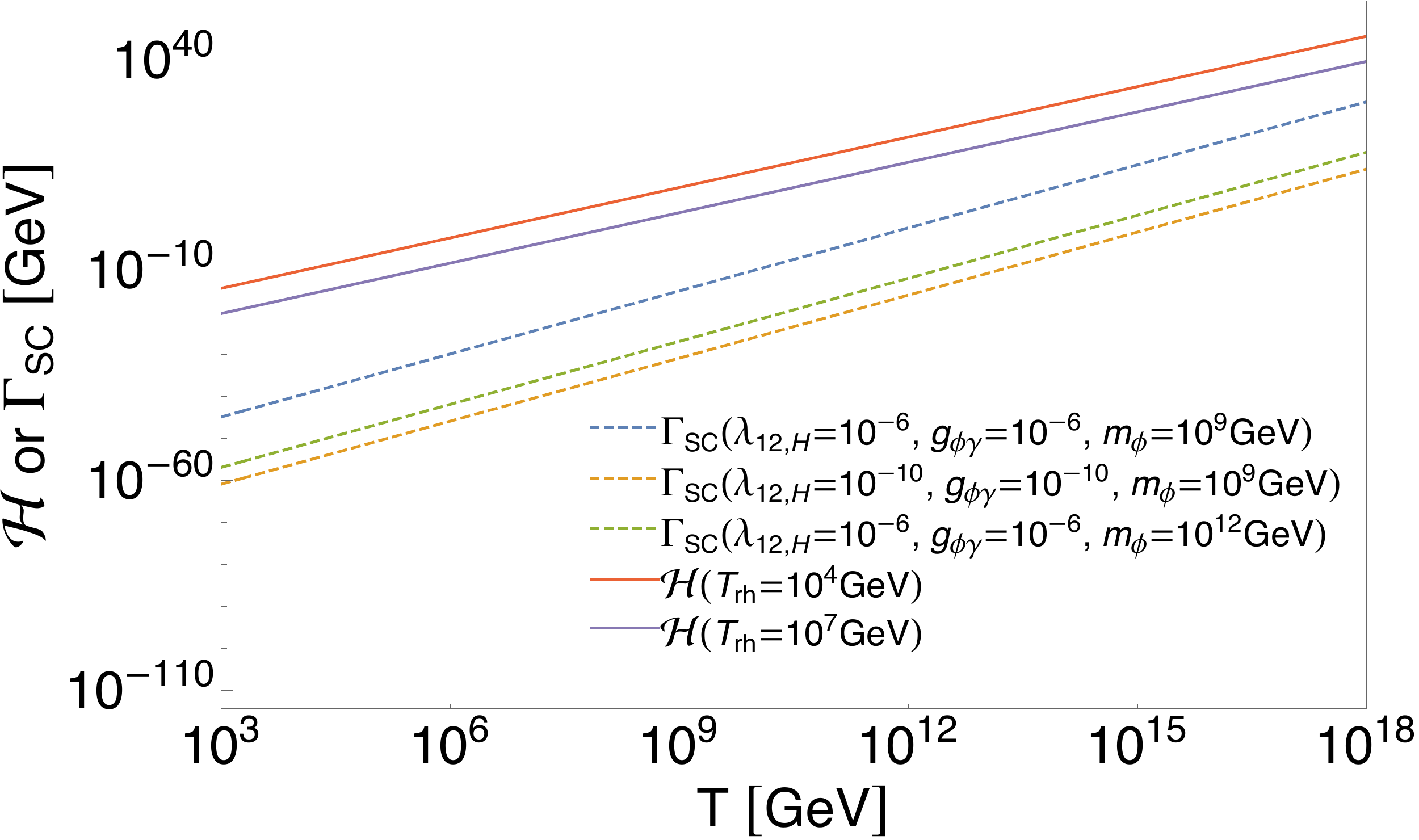}
    \caption{\it \raggedright
    Presentation of representative sets of values for the couplings $(\lambda_{12,H},g_{\phi\ug})$ mentioned in~\cref{Eq:Inflaton decaying to two higgs,Eq:DecayToGaugeBoson}, such that \dr~particles, which are $U(1)$ gauge bosons, cannot achieve thermal equilibrium with the \sm~Higgs via inflaton exchange diagram at the tree level. The solid lines correspond to ${\cal H}$ from~\cref{Eq:Hubble during reheating} for two different values of $\Trh$, while the dashed lines correspond to $\Gamma_{SC}$ from~\cref{eq:gauge-dr--scat-Gamma}, for different values of $(\lambda_{12,H},y_\chi)$ and $m_\phi$.}
    \label{fig:gaugeDR-vs-Hubble}
\end{figure}
\cref{fig:gaugeDR-vs-Hubble} depicts sample values of $(\lambda_{12, H},g_{\phi\ug})$ for which the condition $\Gamma_{SC}< {\cal H}$ is upheld. Solid lines in~\cref{fig:gaugeDR-vs-Hubble} are used to represent ${\cal H}$ parameter derived from~\cref{Eq:Hubble during reheating} for two different values of $\Trh$: $\Trh=10^{4}\GeV$ and $\Trh\sim 10^{7}\GeV$. Moreover, the dashed lines correspond to $\Gamma_{SC}$ for various combinations of values of $(\lambda_{12, H},g_{\phi\ug})$ and $m_\phi$. 
\cref{fig:gaugeDR-vs-Hubble} shows that even with $\Trh\sim 10^7\GeV$, and considering $(\lambda_{12, H}\lsim 10^{-6},g_{\phi\ug}\lsim 10^{-6})$, and $m_\phi\sim 10^9\GeV$, \bsm~\dr~particles which are $U(1)$ gauge bosons, cannot establish thermal equilibrium with \sm~Higgs via 2-to-2 scattering processes with inflaton as the mediator.  
 Reheating scenarios with higher values of $\Trh$, 
 requires smaller values of 
 non-thermal condition for \dr~particles necessitates smaller values of $(\lambda_{12, H},g_{\phi\ug})$, or larger $m_\phi$, or possibly both, to prevent DR particles from achieving thermal equilibrium with \sm~Higgs through inflaton mediation scattering processes.

\end{document}